\documentclass[11pt]{article}

\usepackage{amssymb,amsfonts,amsthm,amsmath,amssymb,amscd,amstext,epsfig}
\usepackage{color}

\def\CPH#1{}
\def\SSP#1{}

\newcommand{\bea}{\begin{eqnarray}}
\newcommand{\eea}{\end{eqnarray}}

\def\eps{{\epsilon}}

\def\be{\begin{equation}}
\def\ee{\end{equation}}

\def\rv{u}
\def\cald{{\mathcal D}}
\def\cala{{\mathcal A}}

\begin{document}

\begin{flushright}
PUPT-2289\\
\end{flushright}

\begin{center}
\vspace{1cm} { \LARGE {\bf The Second Sound of SU(2)}}

\vspace{1.1cm}

Christopher P.~Herzog and Silviu S.~Pufu

\vspace{0.7cm}

{Department of Physics, Princeton University \\
     Princeton, NJ 08544, USA }

\vspace{0.7cm}

{\tt cpherzog@princeton.edu, spufu@princeton.edu} \\

\vspace{1.5cm}

\end{center}

\begin{abstract}
\noindent
Using the AdS/CFT correspondence, we calculate the transport coefficients
of a strongly interacting system with a non-abelian SU(2) global symmetry
near a second order phase transition.  From the behavior of the poles in
the Green's functions near the phase transition, we determine analytically the speed of second sound, the conductivity, and diffusion constants.  We discuss similarities and differences between this and other systems with vector order parameters 
such as $p$-wave superconductors and liquid helium-$3$.
\end{abstract}

\pagebreak


\setcounter{page}{1}
\setcounter{equation}{0}

\section{Introduction}

A new and interesting holographic perspective on the physics of superfluids and superconductors
was provided by refs.\ \cite{Gubser:2008px, Hartnoll:2008vx}.  These papers, which
rely on the AdS/CFT correspondence \cite{Maldacena:1997re, Gubser:1998bc, Witten:1998qj}, 
provide a dual description of 
the superconducting phase transition as the instability of a charged
black hole to develop scalar hair.  Recalling that 
the AdS/CFT correspondence maps a strongly interacting
field theory to a classical gravity description,
this new perspective holds promise for deepening our 
understanding of superconductivity in strongly interacting regimes where
BCS theory \cite{BCS} is inadequate.\footnote{%
 See ref.\ \cite{superreview} for a review of the limits of BCS theory when confronted
 with high temperature superconductivity.
}

This paper takes as its starting point the holographic model proposed in refs. \cite{Gubser:2008zu, GubserPufu, Roberts:2008ns}.  While in refs.\ \cite{Gubser:2008px, Hartnoll:2008vx}  the 
black hole develops scalar hair at the phase transition, in refs.\
 \cite{Gubser:2008zu, GubserPufu, Roberts:2008ns} the black hole develops non-abelian hair.
 More precisely, refs.\ \cite{Gubser:2008px, Hartnoll:2008vx} begin with
gravity plus an abelian gauge field and charged scalar, while
refs.\ \cite{Gubser:2008zu, GubserPufu, Roberts:2008ns} omit the scalar and promote the abelian
field to a non-abelian SU(2) gauge field.   Recall that the AdS/CFT dictionary maps
gauge fields on the gravity side of the duality to global symmetries in the field theory.

We are intentionally 
vague about the distinction between a superfluid and superconductor.
As emphasized in ref.\ \cite{Herzog:2008he}, technically, the field theory dual to the black hole construction
 undergoes a superfluid phase transition, i.e.\
spontaneous symmetry breaking of a global symmetry.    
To interpret this transition as a superconducting phase transition, the global symmetry must be weakly
gauged.  For many questions about superconductivity, the distinction is irrelevant, and from the
current-current two point functions we calculate below, we can extract meaningful conductivities.

This SU(2) system has many desirable features compared to the scalar system.  The scalar system appears to be less universal; one must specify a potential for the scalar, or at the very least a mass term.  
In comparison, up to the strength of the gravitational and gauge couplings, 
the form of the Lagrangian for the SU(2) system is completely specified by gauge invariance.
Also, it appears more straightforward to embed the SU(2) system in string theory or other
UV completions.  (See however ref.\ \cite{Denef:2009tp} for recent progress with the scalar system.)
This embedding can move the SU(2) system out of the toy model realm and give us a
microscopic Lagrangian for the field theory.  Refs.\ \cite{Basu, Kaminski} discuss one possible
embedding where the SU(2) is a global flavor symmetry group for ${\mathcal N}=4$
super Yang-Mills broken to ${\mathcal N}=2$ by the addition of two hypermultiplets.\footnote{%
 The most naive embedding does not actually work.  
 The original AdS/CFT correspondence
 gives us a duality between ${\mathcal N}=4$ super Yang Mills and string theory in the
 curved background $AdS_5 \times S^5$.  This string theory can be approximated at low
 energies by a gauged supergravity in $AdS_5$.  The SU(4) gauge field, which maps to 
 the SU(4) global R-symmetry in the field theory, has an SU(2) subgroup.  Unfortunately,
 supersymmetry constrains the relationship between the gauge field coupling and the
 gravitational coupling in this model, and the gauge field coupling is too weak for a 
 superconducting phase transition to occur \cite{Roberts:2008ns}. 
}  

Our reason for choosing this SU(2) system is very simple:  We can get analytic results near the phase transition.  The differential equations that describe these holographic systems are nonlinear, and
analytic solutions do not appear to be available in most cases.  
The studies mentioned above make extensive use of numerics to see the phase transition, to calculate
the conductivities and critical exponents.  
Analytic results,
for example the low temperature approximation of the conductivity in ref.\ \cite{Hartnoll:2008vx},
are scarce.\footnote{%
See refs.\ \cite{Gubser:2008wz,Maeda:2008ir} for other nice analytic results for this class of models.
}
  Our starting point is the remarkable observation in ref.\ \cite{Basu} that the zero mode
for the phase transition for an SU(2) gauge field in $AdS_5$ has a simple analytic form.  
From this zero mode, we are able to extract a long list of properties near the phase transition:
\begin{enumerate}
\item
The speed of second sound near the phase transition.

\item
That the phase transition is second order.

\item
The conductivity and in particular the residue of the
pole in the imaginary part of the conductivity.

\item
The system satisfies a London type equation that implies a Meissner effect.

\item
A large selection of current-current Green's functions in the hydrodynamic limit, 
and that they satisfy the appropriate non-abelian Ward identities.
\end{enumerate}

To clarify the title of the paper, recall that in a two component fluid, there are typically two propagating
collective modes.  The first mode corresponds to ordinary sound in which the two components move in phase.  The second mode corresponds to second sound in which the two components move out of phase.  Typically, ordinary sound can be produced by pressure oscillations while second sound couples much more strongly to temperature oscillations \cite{LandL}. 

The order parameter for the phase transition in our SU(2) model is the set of non-abelian
global SU(2) currents, $j_a^\mu$.  As pioneered in ref.\ \cite{Gubser:2008zu}, we introduce by hand
a chemical potential in the third isospin direction which induces a charge density, $\langle j_3^t \rangle \neq 0$, that breaks both the global SU(2) symmetry to a U(1) sugroup and also 
Lorentz invariance. There is a superconducting phase transition at a critical temperature $T_c$, below which a current develops orthogonal to the third isospin direction that completely breaks the residual U(1) symmetry and also breaks the remaining rotational symmetry of the system to U(1).
For convenience, we take this current to be in the direction $\langle j^1_x \rangle$, leaving
a rotational symmetry in the $yz$-plane.

The fact that rotational symmetry is broken in the superconducting phase makes the physics of this
model rich and complicated.  Our model appears to be a holographic realization of the type of scenario described from a formal perturbative field theoretic point of view in ref.\ \cite{Buchel:2006aa}.
Transport coefficients such as the speed of second sound and
conductivities will depend on which direction we decide to look.  Such a breaking of rotational invariance is not unheard of in real world materials.  To pick a particularly simple example, a ferromagnet will break rotational symmetry when the spins align. Ref.\ \cite{GubserPufu}
emphasized a possible connection of this SU(2) model 
to a $p$-wave superconductor, where the order parameter for the phase transition is a vector.

Of the real world materials that we considered, superfluid liquid helium-$3$ perhaps comes closest in
approximating the physics of our model.
Liquid helium-$3$ at very low temperatures
is a $p$-wave 
superfluid.  
Two fermionic helium-$3$ atoms pair up to form a loosely bound bosonic 
molecule with weak interaction between
the orbital and spin degrees of freedom of the electrons \cite{VollhardtWolfle}.  
The orbital and spin angular momenta are both equal to one, and the order parameter is often
written $A_{a i}$ where $a$ indexes the spin angular momentum and $i$ the orbital angular momentum,
in surprisingly close analogy with our $j^a_\mu$.  There are many stable phases of superfluid helium-$3$, depending on the pressure, temperature, and applied magnetic field.  The A phases are known to break rotational symmetry.  

Despite plausible similarities between the symmetries of our model and various real world materials, there is one crucial difference.  While the order parameters for these real world materials may have vector or tensor structure, they are not currents, and the signature of the phase transition is not the production of a persistent current.  In contrast, our model has $\langle j_1^x \rangle \neq 0$. 

We begin in Section 2 with a discussion of the SU(2) model and the probe limit.  We choose to work in a limit in which gravity is weak and the non-abelian field does not back react on the metric.  Thus, at heart, in this paper we will be solving the classical SU(2) 
non-abelian Yang-Mills equations in a fixed background spacetime, 
that of a Schwarzschild black hole in $AdS_5$.\footnote{%
 Attempts to solve the full set of coupled equations for a non-abelian black hole go back many years \cite{Bartnik:1988am, Bizon:1990sr}.  See refs.\ \cite{Volkov:1998cc, Winstanley:2008ac} for reviews.
}  

In Section 3, we find a solution to the Yang-Mills equations near the phase transition.  This power series solution in the order parameter and superfluid velocities 
allows us to demonstrate that the phase transition is second order and to calculate the speed of second sound from thermodynamic identities.

In Section 4, we make some formal remarks about the current-current correlation functions for our model.  We discuss the Ward identities that these Green's functions satisfy and some of their discrete symmetries.  We also review how to calculate these two-point functions using the AdS/CFT correspondence.

In Section 5, through a study of fluctuations about our solution near the phase transition, we extract the current-current correlation functions in the hydrodynamic limit.  From the location of the poles, we independently confirm the speed of second sound calculated in Section 3.  We are also able to calculate various damping coefficients and see explicitly that the Green's functions satisfy the non-abelian Ward identities.  In the last part of Section 5, we consider the $\omega \to 0$ and $k \to 0$ limits.  From these limits we extract conductivities and also demonstrate that the system obeys a type of London equation.

\section{The Model}

Consider the following gravitational action for a non-abelian gauge field $F_{AB}^a$ 
with a cosmological constant $\Lambda$:
\be
S = \frac{1}{2 \kappa^2} \int d^{d+1} x \sqrt{-g} \left( R - 2 \Lambda \right) - \frac{1}{4 g^2} \int d^{d+1} x \sqrt{-g} \, F^a_{A B} F^{a AB} \ .
\label{action}
\ee
Our gauge field can be re-expressed in terms of a connection $A_B^a$ as follows:
\be
F_{AB}^a = \partial_A A_B^a - \partial_B A_A^a + {f^a}_{bc} A_A^b A_B^c \ ,
\ee
where ${f^a}_{bc}$ are the structure constants for our Lie algebra ${\mathfrak g}$ with generators $T_a$ such that $[T_a, T_b] = i {f_{ab}}^c T_c$.  We will take ${\mathfrak g} =  {\rm su(2)}$ where $T_a = \sigma_a / 2$, $\sigma_a$ are the Pauli spin matrices, and the structure constants are $f_{abc} = \eps_{abc}$.\footnote{%
The ${\mathfrak g} =  {\rm su(2)}$ indices $a, b,c, \ldots$ are raised and lowered with the Kronecker delta $\delta^a_b$.  The capital indices $A,B,C \ldots$ are raised and lowered with the five dimensional space time metric $g_{AB}$.  We will also shortly introduce Greek indices $\mu,\nu, \ldots$ which will be raised and lowered with the four dimensional Minkowski tensor $\eta^{\mu\nu} = ({-}{+}{+}{+})$.
}

The equations of motion for the gauge field that follow from this action (\ref{action}) are 
$D_A F^{a AB}= 0$ which can be expanded as
\be
\nabla_A F^{a AB} + {f^a}_{bc} A_A^b F^{c AB} = 0 \ .
\label{gaugefieldeom}
\ee
Einstein's equations can be written 
\be
 R_{AB} + \left( \Lambda - \frac{1}{2} R \right) g_{AB} 
= \frac{\kappa^2}{2 g^2} \left( 2
F^a_{C A} {F^{aC}}_B - \frac{1}{2} F^a_{CD} F^{a CD} g_{AB}  \right)  \ .
\ee

A solution to these equations in the case of a negative cosmological constant, $\Lambda= - d (d-1)/2$,
is a $d+1$-dimensional Reissner-Nordstr\"om black hole with anti-de Sitter space asymptotics.\footnote{%
 We set the radius of curvature $L=1$. 
}
The only non-zero component of the vector potential can be taken to be\footnote{%
 The notation $\tilde \rho$ is meant to evoke a charge density.  The actual
 charge density $\rho = -(d-2) \tilde \rho / g^2$.  
}
\be
A_t^3 = \mu + \tilde \rho \rv^{d-2} \ .
\label{Atansat\rv}
\ee
Thus we are using only a U(1) subgroup of the full SU(2) gauge symmetry; this black hole solution
requires only an abelian gauge symmetry.  The line element for this black hole solution has the form 
\be
ds^2 = {1 \over u^2} \left[-f(\rv)dt^2 + d \vec x^2 + \frac{d\rv^2}{f(\rv)} \right] \ 
\ee
where $d \vec{x}^2 = dx^2 + dy^2 + dz^2$ and the warp factor is
\be
f(\rv) = 1 + Q^2 \left(\frac{\rv}{\rv_h}\right)^{2d-2} - (1+Q^2) \left(\frac{\rv}{\rv_h} \right)^d \ ,
\ee
with the charge $Q$ being defined as
\be
Q^2 \equiv \frac{\kappa^2}{g^2} \frac{d-2}{d-1}\tilde \rho^2 \rv_h^{2d-2} \ .
\ee
The horizon is located at $\rv=\rv_h$, and the Hawking temperature is
\be
T_H = \frac{d-(d-2)Q^2}{4 \pi u_h} \ .
\ee

One subtle issue to be addressed is the boundary conditions for $A_t^3$ at the horizon $\rv=\rv_h$ and
at the boundary $\rv=0$ of our asymptotically AdS space.
At the horizon of the black hole, we must work in a local coordinate patch
for the gauge potential such that $A_t^3$ has a well defined norm, 
$|A_t^2 g^{tt}| < \infty$.  Given the form of $g^{tt}$, we actually require that $A_t(\rv_h) = 0$.
Our gauge potential (\ref{Atansat\rv}) is well defined globally, at both the horizon and the boundary, provided 
\be
\tilde \rho = - \mu/\rv_h^{d-2} \ .
\ee

The boundary value of the gauge potential, $A_t^3(0) = \mu$, is interpreted via the AdS/CFT dictionary as a chemical
potential in the dual field theory.  As such, $\mu$
is an external parameter of the field theory, and we should restrict our class of gauge 
transformations to those which do not affect $\mu$.  For example, the class of abelian
gauge transformations  $A_t^3 \to A_t^3 + \partial_t \Lambda$ 
where $\Lambda = c t$ is ruled out by this restriction. 

Refs.\ \cite{Gubser:2008zu,GubserPufu,Roberts:2008ns} made the observation that below a critical
temperature (or alternately above a critical chemical potential), this charged black hole undergoes a
second order phase transition.  
The component of the gauge field $A_x^1$ develops a profile which not only
spontaneously breaks the remaining $U(1)$ of the $SU(2)$ symmetry but also breaks rotational invariance.  These authors considered the case $d=4$ and interpreted the dual 2+1 dimensional
field theory as a $p$-wave thin film superconductor.  More recently Basu et al.\ \cite{Basu} looked at
the $d=5$ case where they interpreted the dual 3+1 dimensional field theory as a pion 
superfluid.\footnote{%
See also \cite{Kaminski} for a 3+1 dimensional system with similar symmetries and qualitative behavior but a more complicated action.
}

In this paper, we shall make two simplifying assumptions.  The first is to take the probe limit, as was
done in \cite{GubserPufu, Roberts:2008ns, Basu}.  In this limit, $\kappa^2/g^2 \to 0$ and the 
gauge field does not back react on the metric.  The metric remains that of an uncharged black hole
in anti-de Sitter space with warp factor
\be
f(u) = 1 - \left(\frac{u}{u_h}\right)^d \ .
\ee
Next, we restrict to the $d=5$ case because of the observation made in ref.\ \cite{Basu} that the zero mode inducing the phase transition has an analytic form.  Given this form, we are able to compute analytically 
many properties of the field theory close to the phase transition including Green's functions, conductivities, diffusion constants, and the speed of second sound.  We show explicitly that the phase transition is second order.

\section{Critical Behavior}

We specialize to $AdS_5$ where we can construct an analytic 
solution to the gauge field equations of motion
close to the phase transition.  We give the solution as the first few terms in a 
power series in three small parameters:
the order parameter $ \epsilon \equiv g^2 \langle j^x_1 \rangle /2$ 
and chemical-potential-like 
objects we call superfluid velocities, $A_x^3(u{=}0) \equiv {\mathcal A}_x^3 = v_\parallel$ 
and $A_y^3(u{=}0) \equiv {\mathcal A}_y^3 = v_\perp$.  Velocity is a bit of a misnomer here
as the objects $v_\parallel$ and $v_\perp$, like the chemical potential $\mu$, have mass dimension
one.  The name is motivated by their canonical conjugacy to the currents $j_3^x$ and $j_3^y$.

From the AdS/CFT dictionary, the currents 
$\langle j_a^\mu \rangle$ and external field strength ${\mathcal A}_\mu^a$ in the field theory
can be determined from the small $u$ expansion of the bulk gauge field $A_\mu^a$:
\be
\label{ADictionary}
A_\mu^a = \cala_\mu^a + \frac{1}{2} g^2 \langle j_\mu^a \rangle u^2  + \ldots \ .
\ee

\subsection{The Background}

We begin with small steps and construct the solution in the limit $v_\perp = v_\parallel = 0$.
In the probe approximation, the equations of motion for the gauge field take the form 
\be
\cald_t {A^3_t} =  \frac{ (A^1_x)^2}{f} A^3_t  
\qquad \mbox{and} \qquad 
\cald_x {A^1_x} = -\frac{(A^3_t)^2}{f^2} A^1_x   \ ,
\label{A3t0eqandA1x0eq}
\ee
where we have defined the linear second order differential operators
\be
\cald_t \equiv \partial_\rv^2 - \frac{1}{\rv} \partial_\rv \qquad \mbox{and}
\qquad
\cald_x = \cald_y \equiv \partial_\rv^2 + \left( \frac{f'}{f} - \frac{1}{\rv} \right) \partial_\rv \ .
\label{cald}
\ee
To keep the equations simple in what follows, 
we choose to put the horizon of the black hole at $u_h=1$.
To restore units, dimensionful quantities such as the chemical potential $\mu$,
frequencies $\omega$,
and wave-vectors $k$ should be replaced with the dimensionless combinations 
$\mu u_h$, $\omega u_h$, and $k u_h$, respectively.

As pointed out by ref.\ \cite{Basu}, when $A_t^3 = 4 (1-\rv^2)$ there is an analytic solution to the second equation of (\ref{A3t0eqandA1x0eq}) that is regular at the horizon,  of the form
\be
\label{ZeroMode}
A_x^1 = \epsilon \frac{\rv^2}{(1+\rv^2)^2} \ .
\ee
From eq.\ (\ref{ADictionary}), 
the meaning of $\epsilon$ in the dual field theory is, up to normalization, that of an expectation value for the non-abelian current $\langle j_x^1 \rangle = 2 \epsilon/g^2$.  The existence of the solution \eqref{ZeroMode} indicates that the superfluid phase transition occurs when $\mu = 4$.\footnote{%
There are in fact a countable set of such zero modes with $\mu = 4k$ where $k$ is a positive integer.
We discuss these higher zero modes in Appendix \ref{app:higher_zero_modes}.  
As the higher zero modes have higher free energy, they should not affect the phase diagram of the system.
}
Given this zero mode, we look for a general solution to eqs.\ (\ref{A3t0eqandA1x0eq}) as a series expansion in $\epsilon$:
\begin{eqnarray}
A_x^1  &=& \epsilon \frac{\rv^2}{(1+\rv^2)^2} + \epsilon^3 w_1 + \epsilon^5 w_2 + {\mathcal O}(\epsilon^7) \ , \\
\label{At3Series}
A_t^3  &=& 4 (1-\rv^2) + \epsilon^2  \phi_1 + \epsilon^4 \phi_2 + {\mathcal O}(\epsilon^6) \ .
\end{eqnarray}
The solution describes the system for $\mu \gtrsim 4$.  
 Our strategy will be to fix the expectation value of $\langle j_x^1 \rangle = 2 \epsilon/ g^2$ 
 but to allow the chemical potential to be corrected order by order: $\mu = 4 + \epsilon^2 \delta \mu_1 + \epsilon^4 \delta \mu_2 + \ldots$.  Thus, in solving the differential equations, we require the boundary condition that the ${\mathcal O}(u^2)$ term in $w_i$ vanish while $\phi_i(0)$ is allowed to be nonzero.

The differential equation governing $ \phi_1$ is
\be
\cald_t \phi_1 = \frac{4\rv^4}{(1+\rv^2)^5} \ ,
\ee
which has the solution
\be
\phi_1 = (1-\rv^2) \delta \mu_1 + \frac{1}{96} \left( 5 \rv^2 - \frac{ 8 \rv^2 ( 1 + 3 \rv^2 + \rv^4)}{(1+\rv^2)^3} \right) \ .
\ee
We applied the boundary condition that $ \phi_1$ vanish at the horizon.  Also, $\delta \mu_1$ corresponds to a  shift of the chemical potential by $\epsilon^2 \delta \mu_1$.  The value of $\delta \mu_1$ is constrained by the solution for $w_1$, as we now see.  The differential equation for $w_1$ is
\be
 w_1 '' - \frac{1+3 \rv^4}{\rv(1-\rv^4)}  w_1 ' + \frac{16}{(1+\rv^2)^2}  w_1 = -\frac{ 8 \rv^2}{(1-\rv^2)(1+\rv^2)^4}  \delta \phi_1 \ .
\ee
We require the boundary conditions that $w_1$ be regular at the horizon and vanish at the boundary ($\rv=0$).  These conditions leave us with the solution
\be
 w_1 = \frac{c \rv^2}{(1+\rv^2)^2} + \frac{\rv^4( 39 \rv^6 - 331 \rv^4 - 819 \rv^2 - 369)}{20{,}160 (1+\rv^2)^5} + \frac{13 \rv^2 \ln(1+\rv^2)}{1680 (1+\rv^2)^2} \ ,
\ee
and the constraint
\be
\delta \mu_1 = \frac{71}{6720} \ .
\ee
The term in $ w_1$ proportional to $c$ is just the zero mode, and, consistent with our strategy,
 we set $c=0$. 

For the free energy calculation we perform below, we also need the next order corrections, $ \phi_2$ and $ w_2$.  The expressions are too cumbersome to repeat here.  The structure
and boundary conditions are analogous to the case of $\phi_1$ and $ w_1$ considered above.

The near boundary expansion of our solution takes the form
\begin{eqnarray}
A_x^1 &=& \epsilon \rv^2 + {\mathcal O}(u^4) \ , \\
\label{GotAt3Expansion}
A_t^3 &=& \left( 4 + \frac{ 71 \epsilon^2}{6720} + \delta \mu_2 \epsilon^4 
+ {\mathcal O}(\epsilon^6)\right) 
 \\
&& - \left( 4 + \frac{ 281 \epsilon^2}{6720} - \left( \frac{1343 - 1365 \ln 2}{2{,}822{,}400} 
-\delta \mu_2 \right)\epsilon^4 + {\mathcal O}(\epsilon^6) \right) \rv^2 
 + {\mathcal O}(u^4) \ , \nonumber
\end{eqnarray}
where 
 \begin{equation}
   \label{Gotmu2}
   \delta \mu_2 = \frac{13 (-4{,}015{,}679 + 5{,}147{,}520 \ln 2)}{75{,}866{,}112{,}000} \ .
 \end{equation}
These expansions match well with numerical solutions that we found close to the transition temperature.

\subsection{Superfluid flow}
\label{sec:superfluidflow}

In this section, we generalize the background above to allow for the possibility of a superfluid
flow.  In terms of the bulk solution,
this generalization requires turning on a constant value of $A_y^3(u{=}0) = v_\perp$ and
$A_x^3(u{=}0) = v_\parallel$ at the 
boundary corresponding to a non-zero superfluid velocity $(v_\parallel, v_\perp,0)$. 
The differential equations describing this background are a modification of
eqs.\ (\ref{A3t0eqandA1x0eq}):
\begin{eqnarray}
u^2 f \cald_\lambda A^1_\lambda &=& g^{\mu \nu} \left(A_\mu^3 A_\nu^3 A_\lambda^1 - A_\mu^1A_\nu^3 A_\lambda^3 \right) \ , \\
u^2 f \cald_\lambda A^3_\lambda &=& g^{\mu \nu} \left( A_\mu^1 A_\nu^1 A_\lambda^3 
- A_\mu^3 A_\nu^1 A_\lambda^1 \right) \ , \\
A_t^1 \partial_u A_t^3- A_t^3 \partial_u A_t^1&=&   f A_x^1 \partial_u A_x^3  - f A_x^3 \partial_u A_x^1 
+ f A_y^1 \partial_u A_y^3 -f A_y^3 \partial_u A_y^1 \ ,
\end{eqnarray}
where we set $A^2_\mu = A^a_z = A^a_u = 0$.  The repeated covariant $\lambda$ indices on 
the left hand side are not to be summed over.
As before, we solve this system in a small $\epsilon$ expansion, but we also add another
small expansion parameter $\delta \sim v_\perp \sim v_\parallel$.  
There is a non-uniformity in the limit $v_\perp \to 0$ and $v_\parallel \to 0$, and we find two
branches of solutions for small values of the superfluid 
velocity.  In the case where $v_\perp > v_\parallel$, we find 
\begin{eqnarray}
A_t^1 &=&  {\mathcal O}(\epsilon^2) \ , \\
A_x^1 &=& \epsilon \frac{u^2}{(1+u^2)^2} - \epsilon (v_\perp^2 + v_\parallel^2) \frac{ u^2 (u^2+4 \ln (1+u^2))}{24 (1+u^2)^2} + \ldots \ , \\
A_y^1 &=& -\epsilon \frac{v_\parallel}{v_\perp} \frac{u^2}{(1+u^2)^2} +
\epsilon (v_\parallel^2 + v_\perp^2) \frac{v_\parallel} {v_\perp} \frac{u^2 (u^2 + 4 \ln (1+u^2))}{24 (1+u^2)^2}
+ \ldots \ , \\
A_t^3 &=& 4(1-u^2) +   \frac{1}{3} (v_\parallel^2 + v_\perp^2) (1-u^2) 
\\
&&
+\epsilon^2 \frac{v_\perp^2 + v_\parallel^2}{v_\perp^2} \frac{ (1-u^2)(71 + 3u^2 - 627 u^4 -279 u^6)}{6720 (1+u^2)^3} 
+ \ldots \ , \nonumber \\
A_x^3 &=&  v_{\parallel} - \epsilon^2 \frac{v_\parallel}{v_\perp}\frac{v_\parallel^2 + v_\perp^2}{v_\perp} \frac{ u^2(3+9u^2 + 4u^4)}{144 (1+u^2)^3} + \ldots  \ , \\
A_y^3 &=&v_\perp   - \epsilon^2 \frac{v_\parallel^2 + v_\perp^2}{v_\perp} \frac{ u^2 (3 + 9u^2+4u^4)}{144 (1+u^2)^3} + \ldots  \ .
\end{eqnarray}
In the case $v_\perp < v_\parallel$, we find
\begin{eqnarray}
A_t^1 &=& \epsilon \frac{v_\perp^2 + v_\parallel^2}{v_\parallel}
 \frac{u^2 (1-u^2)}{4(1+u^2)} + \ldots \ , \\
A_x^1 &=& \epsilon \frac{u^2}{(1+u^2)^2} + \epsilon (v_\perp^2 + v_\parallel^2) \frac{ u^2 (u^2-2 \ln (1+u^2))}{24 (1+u^2)^2} + \ldots \ , \\
A_y^1 &=& \epsilon \frac{v_\perp} {v_\parallel}\frac{u^2}{(1+u^2)^2} +
\epsilon (v_\perp^2 + v_\parallel^2)
\frac{v_\perp}{v_\parallel}  \frac{u^2 (u^2 -2 \ln (1+u^2))}{24 (1+u^2)^2} + \ldots \ , \\
A_t^3 &=& 4(1-u^2) + \frac{1}{6} (v_\parallel^2 + v_\perp^2) (1-u^2) 
\\
&&
+ \epsilon^2 \frac{v_\perp^2 + v_\parallel^2}{v_\parallel^2} \frac{ (1-u^2)(71 + 3u^2 - 627 u^4 -279 u^6)}{6720 (1+u^2)^3} + \ldots \nonumber
\ , \\
A_x^3 &=&  v_{\parallel} - \epsilon^2 \frac{v_\parallel^2 + v_\perp^2}{v_\parallel} \frac{ u^2(3+9u^2 -2u^4)}{288 (1+u^2)^3}  + \ldots  \ , \\
A_y^3 &=&v_\perp   - \epsilon^2 \frac{v_\perp}{v_\parallel}\frac{v_\parallel^2 + v_\perp^2}{v_\parallel} \frac{ u^2 (3 + 9u^2-2u^4)}{288 (1+u^2)^3} + \ldots  \ .
\end{eqnarray}

These solutions can be used to compute the speed of second sound perpendicular and
parallel to the order parameter $A_x^1$.
In a two component fluid, there are typically two propagating collective modes, ordinary and second sound.  In our probe approximation, we see only the superfluid component, and the single
collective motion available to us we call second sound.  From our holographic perspective,
ordinary sound would involve fluctuations of the metric so it is suppressed in the limit $\kappa^2 / g^2 \to 0$.

The speed of second sound, like that of ordinary
sound, can be computed from derivatives of the state variables.  
From ref.\ \cite{Herzog:2008he}, the second sound speed squared in this probe limit should
be
\be
c_2^2 = \left. -\frac{ \partial j / \partial v}{\partial \rho / \partial \mu} \right|_{v=0} \ .
\label{speedsecondsound}
\ee
From eq.\ (\ref{ADictionary}), 
the values of the charge current $j=\langle j_i^x \rangle$ 
and the charge density $\rho= \langle j_3^t \rangle$ can 
be read off from the order $u^2$ pieces of $A_i^3$ and $A_t^3$, respectively.

Because our system is not rotationally symmetric, the speed of second sound will depend on the 
direction of propagation.  Let $c_\perp$ and $c_\parallel$ be the speeds perpendicular and
parallel to the order parameter $A_x^1$, respectively.  The speed $c_\perp$ can be computed
from the background solution $v_\perp > v_\parallel$ while $v_\parallel$ can be computed
from the solution with $v_\parallel > v_\perp$.  In the case $v_\perp>v_\parallel = 0$, we find that 
\begin{eqnarray}
\label{Atsusc}
A_t^3 &=& \mu + \frac{1}{71}(840 - 281 \mu) u^2 + \ldots \ , \\
A_y^3 &=& v_\perp - \frac{140}{71} v_\perp (\mu-4) u^2 + \ldots \ .
\label{Ayperp}
\end{eqnarray}
and hence, up to higher order corrections in $\epsilon$, 
\be
c_\perp^2 \approx \frac{71}{13{,}488} \epsilon^2 \approx  \frac{140}{281} (\mu - 4) \ .
\label{speedsoundperp}
\ee
(We used the fact that $\mu - 4 \approx 71 \epsilon^2 / 6720$, which can be read off from \eqref{GotAt3Expansion}.)  In the case $v_\parallel>v_\perp=0$, at leading order $A_t^3$ remains the same but now we need
\be
A_x^3 = v_\parallel - \frac{70}{71} v_\parallel (\mu-4) u^2 + \ldots \ .
\label{Aypar}
\ee
We find that
\be
c_\parallel^2 \approx \frac{1}{2} \frac{71}{13{,}488}\epsilon^2 \approx \frac{70}{281} (\mu-4)\ .
\label{speedsoundpar}
\ee
We confirm these results for $c_\parallel$ and $c_\perp$ in Sections~\ref{sec:TransverseSound} and~\ref{sec:LongitudinalSound} through an analysis of the hydrodynamic 
poles in the current-current correlation functions.  For numerical results valid when $\epsilon$ is not necessarily small, see Figure~\ref{c2Plot}.

These perturbative solutions in $v_\perp$ and $v_\parallel$ can also be used to analyze
the phase diagram of the system near the critical point $\mu_c = 4$.  At the critical point,
we expect the order parameter to vanish, so $\epsilon=0$.  The value of $A_t^3$ at $u=0$ can
be reinterpreted as the value of the chemical potential.  These two facts give us a relation between
the chemical potential and superfluid velocity along the critical line separating the two phases.
For superfluid flow parallel to the order parameter, we expect
\be
\mu \approx 4 + \frac{1}{6} v_\parallel^2
\ee
while for flow perpendicular to the order parameter, we have instead
\be
\mu \approx 4 + \frac{1}{3} v_\perp^2 \ .
\ee

\subsection{The Free Energy}

We compute the contribution to the free energy from the gauge field term in the on-shell 
action: 
\begin{equation}
\begin{split}
S &=-\frac{1}{4g^2} \int d^5 x \sqrt{-g} F_{AB}^a F^{a AB} \\
&= \frac{ \beta \mbox{Vol}_3 }{2g^2}\int_0^1 \frac{d\rv}{\rv} \left( (\partial_\rv A_t^3)^2 - f(\rv) (\partial_\rv A_x^1)^2 + \frac{1}{f(\rv)}(A_x^1 A_t^3)^2\right) \\ 
&= -\frac{\beta \mbox{Vol}_3 }{2g^2} \left( \int_0^1 \frac{d\rv}{\rv} f(\rv) (\partial_\rv A_x^1)^2 + \left.\frac{1}{\rv} A_t^3 (\partial _\rv A_t^3) \right|_{\rv=0} \right) \ .
\end{split}
\end{equation}

For a background where $A_x^1=0$ and 
\be
A_t^3 = (4 + \delta \mu_1 \epsilon^2 + \delta \mu_2 \epsilon^4) (1-\rv^2) \ ,
\ee
the on-shell action is
\be
S_{\rm vac} = \frac{\beta \mbox{Vol}_3 }{4g^2} \left( 64 + \frac{71}{210} \epsilon^2 + \left( -\frac{51{,}145{,}217}{2{,}370{,}816{,}000} + \frac{4979 }{176{,}400}\ln 2 \right) \epsilon^4 + {\mathcal O}(\epsilon^6) \right)  \ .
\ee
Here $\mbox{Vol}_3$ is the spatial volume of the field theory while $\beta = 1/T$ is the inverse temperature.  For the background where $A_x^1 \neq 0$ has condensed, we find in contrast that
\be
S_{\rm sf} = \frac{ \beta \mbox{Vol}_3 }{4g^2} \left( 64 + \frac{71}{210} \epsilon^2 + \left( - \frac{48{,}014{,}117}{2{,}370{,}816{,}000} + \frac{4979 }{176{,}400}\ln 2 \right) \epsilon^4+ {\mathcal O}(\epsilon^6)  \right)\ .
\ee
The difference in the values of the two on-shell actions is
\be
\beta \Delta P = S_{\rm vac} - S_{\rm sf} = \frac{\beta \mbox{Vol}_3 }{4 g^2} \left(-\frac{71}{53{,}760} \epsilon^4 + {\mathcal O}(\epsilon^6) \right) \ .
\ee
Now $\Delta P$ can be interpreted also as a difference in the free energies because the free energy (in the grand canonical ensemble) is minus the value of the on-shell action.  That $\Delta P < 0$ implies that the free energy of the superfluid phase is smaller and thus the superfluid is stable.

Moreover, from the fact that the free energy difference scales as $\epsilon^4$, 
we see that the phase transition is second order.  For small $\epsilon$, $\epsilon^4 \sim (\mu- \mu_c)^2$.  If we restore dimensions, then $\mu$ should be replaced by $\mu u_h = \mu /\pi T$. 
Thus,
$\epsilon^4 \sim (T_c-T)^2$.  The derivative of $P$ with respect to temperature is
continuous but non-differentiable.

\section{Formal Remarks about Green's Functions}
\label{sec:wardidentities}

A field theory with a non-abelian global symmetry, such as SU(2), has by Noether's Theorem,
a conserved current $j^\mu_a$ which transforms under the adjoint representation of 
this symmetry group.  In this paper, we are interested in Green's functions for this current,
in particular the Fourier transformed retarded current-current correlation functions:
\be
 \label{GmunuDef}
G^{\mu\nu}_{ab} (p) = i \int d^4x \, e^{-i p \cdot x} \langle [ j_a^\mu(x), j_b^\nu(0) ] \rangle \theta(t) \ .
\ee
If the symmetry is non-anomalous, then we can weakly gauge it by coupling the current
to an external gauge field $\cala_\mu^a$.  Gauge invariance then implies that the correlation functions
obey a series of Ward identities.  For the one-point function, the covariant
derivative of the current vanishes:
\be
    0 = \left( \partial_\mu \delta^c_a + {f_{ab}}^c \cala^b_\mu \right) \langle j^\mu_c \rangle
    \,.
\label {D j = 0}
\ee
More usefully for the present discussion, there is also a Ward identity for the retarded two-point function.
We give here the Fourier transformed version:
\be
0=(i p_\mu \delta^c_a + {f_{ab}}^c \cala_\mu^b ) G^{\mu\nu}_{cd}(p) + {f_{ad}}^c \langle j^\nu_c \rangle  \ .
\ee

For our gravitational system beyond the phase transition, the gravitational bulk values of both 
$A_t^3$ and $A_x^1$ are non-zero.  The AdS/CFT dictionary allows us to read 
$\cala_\mu^a$ and $\langle j^\mu_a \rangle$ from the near boundary expansion of $A_\mu^a$
using the relation (\ref{ADictionary}).  For the system under consideration here, of the components of the external gauge field only $\cala_t^3 = \mu$ is non-zero in the field theory.  We have two non-vanishing components of the current, $\langle j^x_1 \rangle$ and $\langle j^t_3 \rangle$.

Below, we compute the Green's functions that describe the response of the system to an external
gauge field in the third isospin direction: $G^{\mu\nu}_{a3}$.   The relevant Ward identities 
componentwise are 
\begin{eqnarray}
0 &=& i p_\mu G^{\mu \nu}_{3a}  - \langle j^x_1 \rangle \delta_{a2} \delta^{\nu x} \ , 
\label{Wardone}
\\
0 &=& i p _\mu G^{\mu \nu}_{23} + \mu G^{t \nu}_{13} + \langle j^x_1 \rangle \delta^{x \nu} \ , 
\label{Wardtwo}
\\
0 &=& i p_\mu G^{\mu \nu}_{13} - \mu G^{t\nu}_{23} \ .
\label{Wardthree}
\end{eqnarray}
Our Green's functions below obey this set of Ward identities.

Another important observation for the Green's functions under consideration is the symmetry
under swapping the indices.  We observe that
\be
G_{ab}^{\mu\nu}(p) = (-1)^{\phi(a,b)} G_{ba}^{\nu\mu}(p) \ ,
\label{PT}
\ee
where $\phi(a,b)$ is equal to $-1$ if either $a=2$ or $b=2$, but not both, and 1 otherwise.
This symmetry follows from the discrete symmetries of the system.  Given that our currents
are even under PT, i.e.\ parity and time reversal, 
if PT were a symmetry of the state, we would expect the Green's
functions to be symmetric under an index swap.  Our state is not symmetric under PT, but
it is symmetric under PT times a ${\mathbb Z}_2$ operation on the 
${\mathfrak su}(2)$ Lie algebra, $\sigma_1 \to - \sigma_1$ and $\sigma_3 \to -\sigma_3$.

\subsection{Computation of two-point functions}

To compute the current-current correlators \eqref{GmunuDef} in the probe approximation, we perturb the background gauge field by sending
 \begin{equation}
   \label{DeltaADef}
   A_A^a \to A_A^a + \delta A_A^a \ .
 \end{equation}
Consequently, the corresponding field strength $F_{AB}^a$ changes to $F_{AB}^a + f_{AB}^a$, with $f_{AB}^a$ given by
 \begin{equation}
   \label{fDef}
   f_{AB}^a = \partial_A \delta A_{B} - \partial_B \delta A_{A}
     + \epsilon^{abc} \delta A_{A}^b A_{B}^c + \epsilon^{abc} A_{A}^b \delta A_{B}^c \ .
 \end{equation}
From \eqref{action}, one can see that the quadratic action for $\delta A_A^a$ is
 \begin{equation}
   \label{QuadAction}
   S_2 = -{1 \over 4 g^2} \int d^5 x\, \sqrt{-g} f_{AB}^a f^{AB a} \ ,
  \end{equation}
which gives the linearized equations of motion
 \begin{equation}
   \label{DeltaAeoms}
   \nabla^A f_{AB}^a + \epsilon^{abc} A^{A b} f_{AB}^c 
     + \epsilon^{abc} \delta A^{A b} F_{AB}^c= 0 \ .
  \end{equation}
The quadratic action \eqref{QuadAction} is in fact not well defined because the integrand diverges 
as $\ln u$ at small $u$ as we will see.  We will regulate this divergence using holographic renormalization \cite{Skenderis:2002wp}.

 For definiteness, we will only analyze the case where the background gauge field doesn't depend on $t$ or $\vec{x}$ and where its radial components $A_u^a$ vanish.  We choose a similar gauge for the perturbations by requiring $\delta A_u^a = 0$.    Equations \eqref{DeltaAeoms} can be solved approximately in the limit of small $u$.  An appropriate series expansion in this limit is
 \begin{equation}
   \label{DeltaABdy}
   \delta A_\mu^a (t, \vec{x}, u) = \alpha^a_\mu(t, \vec{x}) + \tilde \alpha^a_\mu(t, \vec{x}) u^2 \ln u 
     + \beta^a_\mu(t, \vec{x}) u^2 + \ldots \ .
 \end{equation}
 for some vector-valued functions $\alpha(t, \vec{x})$, $\tilde \alpha(t, \vec{x})$, $\beta(t, \vec{x})$, etc.  The values of $\alpha$ and $\beta$ are the only ones that can be specified independently;  all the other functions appearing in this expansion, namely $\tilde \alpha$ and higher order corrections, can be expressed in terms of $\alpha$ and $\beta$.  
 
Plugging \eqref{DeltaABdy} into \eqref{DeltaAeoms} and looking at the term with the lowest power of $u$ in the equation with $B = \nu$, one finds a relation between $\tilde \alpha_\nu^a$ and $\alpha_\nu^a$:
  \begin{equation}
   \label{GotAlphaTilde}
   \tilde \alpha_\nu^a = -{1 \over 2} \left[ \partial^\mu f_{\mu\nu}^a + \epsilon^{abc} A^{\mu b} f_{\mu\nu}^c 
     + \epsilon^{abc} \delta A^{\mu b} F_{\mu\nu}^c\right] \bigg|_{u = 0} \ .
  \end{equation}

Upon integration by parts in \eqref{QuadAction}, the unregularized on-shell quadratic action can be written as
  \begin{equation}
   \label{SOnShell}
   S_2^{\rm on-shell} = {1 \over 2g^2} \int d^4 x \, {1\over u} (\delta A^{\nu a})
   (\partial_u \delta A_{\nu}^a) \bigg|_{u = 1/\Lambda} \ .
 \end{equation}
The divergence that arises as one takes $\Lambda \to \infty$ comes from the $\partial_u \delta A_\nu^a$ term 
whose most divergent piece goes like $u \ln u$ at small $u$. This divergence can be regulated by adding the counterterm
\begin{equation}
   \label{Counterterm}
   S_{\rm ct} = -{\ln \Lambda \over 2 g^2} \int d^4 x \, \delta A^{\nu a} 
     \left[ \partial^\mu f_{\mu\nu}^a + \epsilon^{abc} A^{\mu b} f_{\mu\nu}^c 
     + \epsilon^{abc} \delta A^{\mu b} F_{\mu\nu}^c\right] \bigg|_{u = 1/\Lambda} \ .
 \end{equation}
Note that this counterterm depends only on the values of the gauge field on the surface $u = 1/\Lambda$ and on its derivatives along this surface, as required by holographic renormalization.

As a side note, a simpler formula for $\tilde \alpha_\nu^a$ can be found if only $A_t^3$ and $\alpha_\nu^3$ approach non-zero values at the boundary of AdS\@.  In this case, only $\tilde \alpha_\nu^3$ is non-zero and is given by
 \begin{equation}
   \label{TildeAlphaSimple}
   \tilde \alpha_\nu^3 = -{1\over 2} \left( 
   \partial_\mu \partial^\mu \alpha_\nu^a  - \partial^\mu \partial_\nu\alpha_\mu^a 
   \right)
   \ .
 \end{equation}
Assuming that $\alpha_\nu^a(t, \vec{x}) = \alpha_\nu^a e^{-i (\omega t - \vec{p} \cdot \vec{x})}$ then 
 \begin{equation}
   \label{TildeAlpha3}
   \tilde \alpha_\nu^3 = {1\over 2} (\vec{p}^2 - \omega^2) \alpha_\nu^3 
     - {1\over 2} p_\nu \left(\omega \alpha_t^3 +
      p_x \alpha_x^3 + p_y \alpha_y^3 + p_z \alpha_z^3 \right)  \ .
 \end{equation}

To compute the Fourier transformed two-point function, we first Fourier transform the regulated on-shell action
\be
S_2^{\rm on-shell} = \frac{1}{g^2} \int \frac{d^4p}{(2\pi)^4} \,  \alpha^\mu_a(-p)
 (\beta_\mu^a(p) + c \tilde \alpha_\mu^a(p)) \ ,
\ee
where $c$ is an arbitrary constant introduced by the regularization procedure.
Although such an action is not a generating functional for the retarded Green's function, using the
procedure outlined by Son and Starinets \cite{Son:2002sd}, we can identify the retarded Green's function as\footnote{%
 For a more precise discussion of how to derive these Green's functions from an action principle and generating functional, see ref.\ \cite{Herzog:2002pc}. See also ref.\ \cite{Skenderis:2008dg} and the discussion in Appendix C of ref.~\cite{Gubser:2008sz}.
}
\be
G^{a \nu}_{\mu b} (p) = \frac{2}{g^2} 
  {\partial \left[ \beta_\mu^a(p) + c \tilde \alpha_\mu^a(p) \right]
  \over \partial \alpha_\nu^b(p)} 
   \ .
\ee
The linear response of a system to a perturbation $\alpha_\nu^b(p)$ is then a current density
of the form
\be
\langle j^a_\mu(p) \rangle = G^{a \nu}_{\mu b}(p) \alpha_\nu^b (p) \ .
\label{twopointformula}
\ee
For most physical questions, the ambiguity in the choice of $c$ should be irrelevant. More precisely,  one can see from \eqref{GotAlphaTilde} that schematically $\tilde \alpha = \partial \partial \alpha + \partial \alpha + \alpha$, so $G_{\mu b}^{a\nu}(p)$ is ambiguous up to an additive term analytic in $p$.  Its Fourier transform $G_{\mu b}^{a\nu}(x)$ is ambiguous up to an additive term of the from $c_1 \delta^4(x) + c_2^\lambda \partial_\lambda \delta^4(x) + c_3^{\lambda \rho} \partial_\lambda \partial_\rho \delta^4(x)$, where $c_1$, $c_2^{\lambda}$, and $c_3^{\lambda\rho}$ are constants that depend on the particular Green's function we are computing.  Since in position space equation \eqref{twopointformula} reads
\be
\langle j^\mu_a(x) \rangle = \int d^4x' \, G^{\mu\nu}_{ab}(x-x') \alpha_\nu^b(x')  \ ,
\ee
it follows that the ambiguity in the choice of $c$ does not affect the result of $\langle j^\mu_a(x)\rangle$ if $\alpha_\nu^b(x) = 0$.  In particular, the late-time, large-distance response of the system to localized sources is not affected by this ambiguity.  There are many subtleties in these calculations.

\section{Fluctuations}

To calculate the Fourier transformed retarded current-current correlation functions, we need to study fluctuations of the SU(2) gauge fields $A^a_\mu(x)$ in our black hole background.  

In the superfluid phase, the expectation value of the order parameter $A_x^1 \neq 0$
breaks rotational symmetry and makes our task richer and more complicated than in the rotationally symmetric case where only $A_t^3 \neq 0$.  
In the rotationally symmetric case, it would be enough to consider a fluctuation with a time and 
space dependence of the form $e^{-i \omega t + i k x}$.  Given the breaking of rotational symmetry,
we should in principle consider a more general dependence where we allow for motion both
parallel and transverse to the order parameter: $e^{- i \omega t+ i k_x x + i k_y y}$.  Because of the complexity of the full result, 
we shall
not present a full accounting of all the Green's functions here.  Instead we will content ourselves by
studying various informative limits where either $k_x=0$ or $k_y=0$.

We make a few other additional simplifying restrictions.  Following in the footsteps of refs.\
\cite{GubserPufu, Roberts:2008ns} where the third isospin direction was interpreted as the 
U(1) of electricity and magnetism, we will consider Green's functions where at least one of the SU(2)
isospin indices is equal to three.  In other words, we are interested in the linear response of the system
to external electric and magnetic fields.  

The last simplifying restriction is to limit our study to the hydrodynamic regime, where the 
order parameter, the frequency, and the wave-vector are small
compared to the temperature.  In our dimensionless notation, $\epsilon, k, \omega \ll 1$.  
It is only in this limit that we have analytic results although
it is straightforward to calculate the Green's functions numerically beyond this regime.

We work out the Green's functions in five cases.  The first and simplest case, for which
we give the most detailed description of the calculation, is for
a fluctuation transverse to the order parameter and a wave vector transverse to both
the order parameter and the polarization of the fluctuation.  We call this fluctuation the
pure transverse mode.
We next consider fluctuations that correspond to a second 
sound mode in two different limits, one where the sound is propagating parallel to the order
parameter and one where the sound is propagating transverse.  These two sets of 
fluctuations give us independent confirmation of the speeds of second sound computed  in Section~\ref{sec:superfluidflow} from thermodynamics.  Finally we consider fluctuations that correspond to a diffusive mode, again in two different limits, one where the diffusion is parallel to the order parameter, one in which the diffusion is transverse.
(In Appendix \ref{sec:generalcase}, we attempt to give a picture of the general case and how the different limits of these Green's functions fit together.)
In Section \ref{sec:conductivities}, we discuss conductivities and the London equations.

In what follows, to avoid cumbersome indices, we define new variables for the background
values of the gauge field:
\be
A_x^1 \equiv W \qquad \mbox{and} \qquad A_t^3 \equiv \Phi \ .
\ee

\subsection{Pure transverse mode}

The pure transverse mode is described by fluctuations of the field $A_y^3$ with
only $z$ spatial dependence.  We decompose
the fluctuations into Fourier modes:
\be
\delta A_y^3(u,t,z) = a_y(u) e^{-i \omega t + i k z} \ .
\ee
These modes
transverse to the order parameter $A_x^1$ decouple from the other fluctuations of the gauge field and
are governed by the differential equation: 
\be
\cald_y a_y = \frac{ (k^2 + W^2) f - \omega^2}{f^2} a_y \ ,
\label{ptshear}
\ee
where $\cald_y$ was defined in eq.\ (\ref{cald}).

Near the horizon $u=1$, we find that 
$
a_y \sim (1-\rv)^{\pm i \omega/4}
$
satisfies either ingoing or outgoing plane wave type boundary conditions. 
Consistent with the presence of an event horizon, it is natural to choose ingoing boundary conditions
(the minus sign in the exponent).  This choice leads to retarded, as opposed to advanced, Green's functions in the dual field theory \cite{Son:2002sd}.  At the boundary $u=0$ of AdS, we would like the freedom to set
$a_y(0) = a_{y0}$ to some arbitrary value of our choosing, 
corresponding to perturbing the dual field theory by
a small external field strength.  These two boundary conditions along with the differential equation uniquely specify the functional form of $a_y$.

While an analytic solution to eq.\ (\ref{ptshear}) does not appear to be available, one can easily solve
this equation in the limit of small $\omega$, $k$, and $\epsilon$. 
We can write the solution for $a_y$, valid to order $\epsilon^2 k$, $\epsilon^2 \omega$, $k^2$, and $\omega^2$, in the form
\be
\label{ayshear}
a_y = 
a_{y0}  \left( \frac{1-\rv^2}{1+\rv^2} \right)^{-i \omega/ 4}
\left(
1+ \epsilon^2 a_{y\epsilon} + \epsilon^2 \omega a_{y\omega \epsilon}  
+ k^2 a_{y k}
+ \omega^2 a_{y \omega} +
\ldots
\right) \ .
\ee
We find 
\begin{eqnarray}
a_{y\epsilon} &=& - \frac{\rv^2 (3 + 9\rv^2 + 4 \rv^4)}{144 (1+\rv^2)^3} \ , \\
a_{y\omega\epsilon} &=& - \frac{ i u^2 (12 + 27 u^2 + 13 u^4)}{864 (1+u^2)^3} \ , \\
a_{yk} &=& \frac{1}{8} \left( 2 \ln (\rv)  \ln \left( \frac{1+\rv^2}{1-\rv^2} \right) + 
\mbox{Li}_2(-\rv^2) - \mbox{Li}_2(\rv^2) \right) \ .
\end{eqnarray}
The expression for $a_{y\omega}$ is too cumbersome to give here.
Near the boundary, this solution (\ref{ayshear}) has the expansion
\be
a_y = a_{y0} + a_{y0} \left(\frac{i\omega}{2} - \frac{\epsilon^2}{48} - \frac{i \omega \epsilon^2}{72}
 -\frac{\omega^2 \ln 2}{4} + \frac{1}{2} (\omega^2-k^2) \left(\frac{1}{2} -\ln (\rv)\right)
\right) \rv^2 + \ldots
\ee
From this near boundary expansion and eq.\ (\ref{twopointformula}), we can calculate the two-point function for the current in the hydrodynamic limit: 
\be
G^{yy}_{33}(\omega,k) = \frac{2}{g^2} 
\left( \frac{i\omega}{2} - \frac{\epsilon^2}{48} - \frac{i \omega \epsilon^2}{72}
 -\frac{\omega^2 \ln 2}{4} + (\omega^2-k^2) c 
\right) + \ldots .
\label{puretransverseG}
\ee
Note that the counter-term ambiguity, proportional to an arbitrary constant $c$,
 is of the form predicted in eq.\ (\ref{TildeAlpha3}).

\subsection{Transverse sound fluctuations}
\label{sec:TransverseSound}

In general, second sound modes are expected to produce poles in the density-density correlation
function. 
We thus need to consider fluctuations in the conjugate field $A_t^3$.
 If we consider sound modes moving transverse to the order parameter, we can take
the fluctuations to have a $y$ dependence but no $x$ dependence.  
The self-consistent set of fluctuations to consider that couple to $\delta A_t^3(u,t,y)$ are
\begin{eqnarray}
\delta A_t^3(u,t,y) &=& a_t^3(u) e^{-i \omega t + i k y} \ , \nonumber \\
\delta A_y^3(u,t,y) &=& a_y^3(u) e^{-i \omega t + i k y} \ , \\
\delta A_x^a(u,t,y) &=& a_x^a(u) e^{-i \omega t + i k y} \ , \nonumber
\end{eqnarray}
where $a = 1,2$.

The four fluctuations
satisfy four second order ordinary differential equations and one first order constraint:
\begin{eqnarray}
\label{ax1eqperp}
\cald_x a_x^1 &=&   
\left( \frac{-\omega^2-\Phi^2 + k^2 f}{f^2}  \right) a_x^1 
+ \frac{2\Phi(  i \omega \Phi  a_x^2 - W a_t^3 )}{f^2} \ , \\
\label{ax2perp}
\cald_x a_x^2  &=&  
\left( \frac{-\omega^2-\Phi^2 + k^2 f}{f^2}  \right) a_x^2 
 - \frac{2 i \omega \Phi  a_x^1 -
i W ( \omega a_t^3 + k f a_y^3)}{f^2} \ , \\
\label{ay3perp}
\cald_y a_y^3  &=& 
 \frac{-\omega^2 + W^2 f}{f^2}a_y^3 - \frac{k \omega}{f^2} a_t^3 + \frac{i k W}{f} a_x^2 \ , \\
\label{at3perp}
\cald_t a_t^3  &=&  -\frac{ k^2 + W^2}{f} a_t^3 
+ \frac{ \omega k }{f} a_x^3 + \frac{2 W \Phi}{f} a_x^1- \frac{ i\omega}{f} a_x^2 \ , \\
0 &=&\frac{i  \omega}{f}  \partial_\rv a_t^3 + i k  \partial_\rv a_y^3 +  W \partial_\rv a_x^2 -  (\partial_\rv W) a_x^2 \ ,
\label{constraintperp}
\end{eqnarray}
where $\cald_t$, $\cald_x$, and $\cald_y$ were defined in eq.\ (\ref{cald}).  We checked that the derivative of the constraint equation (\ref{constraintperp}) with respect to $u$ is a linear combination of all five differential equations \eqref{ax1eqperp}--\eqref{constraintperp}. Thus if a solution of the first four differential equations satisfies the constraint for some $\rv$, 
it will satisfy the constraint equation at all $\rv$.  

There are seven integration constants associated with this linear system \eqref{ax1eqperp}--\eqref{constraintperp}.  If we look at the horizon of the black hole at $\rv = 1$, we find seven different kinds of behavior.  There exist six solutions that have plane wave behavior for $a_x^1$, $a_x^2$, and $a_y^3$ near the horizon of the form
\be
(1-u)^{\pm i \omega / 4} \ .
\ee
There is also a pure gauge solution, 
\be
a_t^3 = - i \omega  \ , \qquad
a_x^3 = i k  \ , \qquad
a_x^2 = -W \ .
\ee
As in the pure transverse case, we choose pure ingoing boundary conditions corresponding to 
$(1-u)^{-i \omega / 4}$ behavior.  
At the boundary $u=0$ of our asymptotically AdS space, we would like to be able to perturb the system with arbitrary boundary values of $a_t^3$ and $a_y^3$ but
set the ``unphysical'' components of the gauge field $a_x^1$ and $a_x^2$ to zero.  These are four
constraints and we have only three ingoing solutions.  Thus we will also need to make use of the pure gauge
solution to enforce our $u=0$ boundary conditions.

We solved the system perturbatively in $\omega$, $k$, and $\epsilon$.  
We present the results here in the limit where $\omega \sim k^2 \sim \epsilon^2$. The near boundary expansion ($u=0$) of the solution takes the form 
\begin{eqnarray}
a_x^1 &=&  - \frac{  (a_{t0} k + a_{y0} \omega)}{{\mathcal P}}  70 k \epsilon 
\left(48 k^2 + 3 \epsilon^2 - 248 i \omega \right)
u^2 + \ldots
\ , \\
a_x^2 &=& - \frac{a_{t0} k + a_{y0} \omega}{{\mathcal P}} \frac{i \omega \epsilon}{k}
\left( 
21{,}840 k^2 + 843 \epsilon^2 - 72{,}800 i \omega
\right)  u^2 + a_{y0} \frac{i \epsilon}{k} u^2 + \ldots \ , \\
a_y^3 &=& a_{y0}
 - \frac{ (a_{t0} k + a_{y0}\omega) }{{\mathcal P}} \omega \Bigl(
1120 k^4 + 3 k^2 (117 \epsilon^2 - 1120 i \omega)  \\
&&
\qquad \qquad \qquad 
 + \frac{1}{48}(\epsilon^2 - 24i \omega) (843 \epsilon^2 - 72{,}800 i \omega)
 \Bigr) u^2
 + \ldots
\ , \nonumber \\
a_t^3 &=& a_{t0} + 
\frac{(a_{t0} k + a_{y0} \omega)}{{\mathcal P}} k
 \Bigl(
1120 k^4  + 3k^2 (117 \epsilon^2 -1120 i \omega)
\label{at3perpsound} \\
&&
\qquad \qquad \qquad 
 +\frac{1}{48}(\epsilon^2 - 24i \omega) (843 \epsilon^2 - 72{,}800 i \omega) 
 \Bigr) 
u^2 
+ \ldots
\ . \nonumber 
\end{eqnarray}
Note, the expression $(a_{t0} k + a_{y0} \omega)$ is not homogeneous in our scaling limit. 
We have included the leading corrections proportional to $a_{t0}$ and $a_{y0}$.
There are terms in the expansion proportional to $u^2 \ln u$ but they are subleading 
in $\omega$, $k$, and $\epsilon$.

The pole in this limit takes the form
\begin{eqnarray}
{\mathcal P} &=& - 72{,}800 i \omega^3 + \left( 43{,}120 k^2 + 843 \epsilon^2 \right) \omega^2
+ \frac{7i}{6} (4800 k^4 + 553 k^2 \epsilon^2) \omega + \nonumber \\
&& - \frac{71k^2 \epsilon^4}{16} - 141 k^4 \epsilon^2 -1120 k^6 + \ldots \ .
\end{eqnarray}
Let us study this cubic polynomial in $\omega$ in two different limits.  First,
if $k \ll \epsilon$, we find three poles with the asymptotic form
\begin{eqnarray}
\omega &=& \pm \sqrt{\frac{71}{13{,}488}} \epsilon k - \frac{147{,}217 i k^2}{947{,}532} + \ldots \ , \\
\omega &=& - \frac{843 i \epsilon^2}{72{,}800} - \frac{4{,}335{,}443 ik^2}{15{,}397{,}395} + \ldots \ .
\label{thirdpole}
\end{eqnarray}
The first two poles are propagating modes that we identify with second sound.  Indeed, the speed of second sound agrees with the earlier result (\ref{speedsoundperp}) from Section \ref{sec:superfluidflow}.  
The position of the third pole in this
limit is determined mostly by the size of the order parameter $\epsilon$ and so we associate it with the
zero mode that causes the phase transition from the superfluid phase back to the normal phase.

In the opposite limit, $k \gg \epsilon$, where the order parameter is small, the behavior should be close to that of the normal fluid.  In this limit, we find
\begin{eqnarray}
\omega &=& \left(  \frac{\pm 11- 3i}{65}  \right) k^2 + \left(  \frac{\pm 260{,}803 -131{,}519i}{26{,}644{,}800}  \right) \epsilon^2 + \ldots
\label{zmodetsound}
 \\
\omega &=& - \frac{ik^2}{2} - \frac{5i \epsilon^2}{2928} + \ldots \ .
\end{eqnarray}
The first two poles are associated with the zero modes that cause the phase transition from the normal phase to the superfluid phase and were discussed in ref.\ \cite{GubserPufu} while the third
pole is associated to the diffusive mode of our conserved charge density.  
Indeed, the location of this diffusive 
pole is determined by the dynamics of the normal phase and was 
calculated, without the order $\epsilon^2$ correction, long ago in ref.\ \cite{SSP}.
As we vary $\epsilon$ and
$k$ the number of poles cannot change.  The two zero mode poles evolve into the sound poles
of the previous limit while the diffusive pole becomes the zero mode pole of the previous limit.  

From these small $u$ expansions, we can read off the eight Green's functions
$G^{xt}_{13}$, $G^{xy}_{13}$, $G^{xt}_{23}$, $G^{xy}_{23}$, $G^{yy}_{33}$, 
$G^{yt}_{33}$, $G^{ty}_{33}$, and $G^{tt}_{33}$.  From the discrete symmetries (\ref{PT}), we
can also read off four more Green's functions with the indices swapped.  Note
the prefactor $a_{t0} k + a_{y0} \omega$ in the small $u$ expansion.  This structure
is necessary to satisfy the Ward identities (\ref{Wardone}).

As a further check, we consider a particular static limit of the density-density correlation function.
From eqs.\ (\ref{twopointformula}) and (\ref{at3perpsound}), we can read off the Green's function,
\be
G^{tt}_{33} = - \frac{2}{g^2} \frac{k^2}{{\mathcal P}}
 \left(1120 k^4 + 3k^2(117 \epsilon^2 - 1120 i \omega) + \frac{1}{48} (\epsilon^2 - 24 i \omega) (843 \epsilon^2 - 72{,}800 i \omega) \right) \ .
\ee
We are interested in the long wave-length limit of this Green's function:
\be
\lim_{k \to 0} G^{tt}_{33} (0, k) = \frac{2}{g^2} \frac{281}{71} \ .
\label{chi}
\ee
This long wave-length limit is equal to a thermodynamic susceptibility, 
\be
\lim_{k \to 0} G^{tt}_{33} (0, k) =  \frac{\partial^2 P}{\partial \mu^2} =  \frac{\partial \rho}{\partial \mu} \ .
\ee
Given this relation, we see that eq.\ (\ref{Atsusc}) agrees with eq.\ (\ref{chi}).

\subsection{Longitudinal sound fluctuations}
\label{sec:LongitudinalSound}

Longitudinal sound modes correspond to the case where the fluctuations in $A_t^3$ depend only on $x$.  A self-consistent set of perturbations in this case is given by
\begin{eqnarray}
\delta A_t^a(u,t,x) &=& a_t^a(u) e^{-i \omega t + i k x} \ , \nonumber \\
\delta A_x^b(u,t,x) &=& a_x^b(u) e^{-i \omega t + i k x} \ ,
\end{eqnarray}
where $a, b = 1,2,3$.   These fields satisfy the following six second order equations and three constraints:
\begin{eqnarray}
\label{at1par}
\cald_t a_t^1 &=& \frac{1}{f} \left(-W \Phi  a_x^3-i k \Phi  a_x^2+k \omega  a_x^1+ k^2 a_t^1 \right)
\ , \\
\label{at2par}
\cald_t a_t^2 &=& \frac{1}{f} \left( 2 i k W a_t^3  + \left (k^2 + 
      W^2 \right) a_t^2 + i W \omega  a_x^3 + 
k \omega  a_x^2 + i k \Phi  a_x^1 \right)
\ , \\
\label{at3par}
\cald_t a_t^3 &=& \frac{1}{f} \left( \left(k^2+W^2\right) a_t^3-2 i k W a_t^2+k \omega  a_x^3-i W \omega  a_x^2+2 W \Phi  a_x^1 \right)
\ , \\
\label{ax1par}
\cald_x a_x^1 &=& \frac{1}{f^2} \left( -2 W \Phi  a_t^3 + i k \Phi  a_t^2 - k \omega  a_t^1 + 
 2 i \Phi  \omega  a_x^2 - \left (\Phi ^2 + \omega ^2 \right) a_x^1 \right)
\ , \\
\label{ax2par}
\cald_x a_x^2 &=& \frac{1}{f^2} \left( -i W \omega  a_t^3 - k \omega  a_t^2 - 
 i k \Phi  a_t^1 - \left (\Phi ^2 + \omega ^2 \right) a_x^2 - 
 2 i \Phi  \omega  a_x^1
 \right)
\ , \\ 
\label{ax3par}
\cald_x a_x^3 &=& \frac{1}{f^2} \left( -k \omega  a_t^3 + i W \omega  a_t^2 + W \Phi  a_t^1 - \omega ^2 a_x^3 \right)
\ , \\
\label{constraint1par}
0 &=& -\Phi ' a_t^2+i \omega  \partial_u a_t^1+\Phi  \partial_u a_t^2+i f k \partial_u a_x^1
\  , \\
\label{constraint2par}
0 &=& f W' a_x^3+ \Phi ' a_t^1 -\Phi  \partial_u a_t^1+i \omega  \partial_u a_t^2+i f k \partial_u a_x^2-f W \partial_u a_x^3
\ , \\ 
\label{constraint3par}
0 &=& -f W' a_x^2+i \omega  \partial_u a_t^3+f W \partial_u a_x^2+i f k \partial_u a_x^3
\ ,
\end{eqnarray}
with $\cald_t$ and $\cald_x$ as defined in \eqref{cald}.  Again, the three constraint equations are consistent with the second order equations in the 
sense that if they hold at some $u$, they hold at all $u$.

The system \eqref{at1par}--\eqref{constraint3par} has nine integration constants.  The nine possible behaviors at the horizon are of two types:  six plane wave solutions for $a_x^1$, $a_x^2$, and $a_x^3$ that behave as
  \be
   (1-u)^{\pm i \omega / 4} 
  \ee
close to $u = 1$, and three pure gauge solutions given by
  \begin{align}
    a_t^1 &= - i \omega \alpha^1 - \Phi \alpha^2 \ , \qquad
    a_t^2 = - i \omega \alpha^2 + \Phi \alpha^1 \ , \qquad
    a_t^3 = - i \omega \alpha^3 \ , \nonumber\\
    a_x^1 &= i k \alpha^1 \ , \qquad
    a_x^2 = i k \alpha^2 - W \alpha^3 \ , \qquad
    a_x^3 = W \alpha^2 + i k \alpha^3 \ ,
  \end{align}
where $\alpha^a$ are arbitrary constants.  As in the previous sections, we require no outgoing modes at the horizon, which amounts to specifying three of the nine integration constants.  The other six integration constants are specified in terms of the values of the fields at $u = 0$.  In order to examine fluctuations in $a_t^3$ and $a_x^3$, we set their boundary values to $a_{t0}$ and $a_{x0}$, respectively, and the boundary values of the other four fields to zero.

Solving the system \eqref{at1par}--\eqref{constraint3par} perturbatively in $\omega$, $k$, and $\epsilon$ under the scaling assumption $\omega \sim k^2 \sim \epsilon^2$, we find that the boundary behavior of the fluctuations is
 \begin{eqnarray}
       a_t^1 &=& 
   \frac{a_{t0} k + a_{x0} \omega}{{\mathcal P}}
   \frac{\epsilon \omega}{4} (20{,}160 k^2 + 843 \epsilon^2 - 72{,}800 i \omega)u^2 + \ldots
       \ , \\
    a_t^2 &=&
     \frac{a_{t0} k + a_{x0} \omega}{{\mathcal P}}
    \frac{35 k^2 \epsilon}{4}(48 i k^2 + 3 i \epsilon^2 + 320 \omega)u^2 + \ldots
       \ , \\
     a_x^1 &=&
      -\frac{a_{t0} k + a_{x0} \omega}{{\mathcal P}} 
    35 k \epsilon (48 k^2 + 3 \epsilon^2 - 320 i \omega)  u^2 + \ldots
       \ , \\
     a_x^2 &=&
    - \frac{a_{t0} k + a_{x0} \omega}{{\mathcal P}} \frac{i\epsilon \omega}{k}
     \bigl(20{,}160 k^2 + 843 \epsilon^2 - 72{,}800 i \omega \bigr) u^2  + a_{x0} \frac{i\epsilon }{k} u^2
     + \ldots \ ,
     \\
    a_t^3 &=& a_{t0} + 
     \frac{a_{t0} k + a_{x0} \omega}{{\mathcal P}}
    \frac{k}{96}
    \bigl( 26{,}880 k^4 + 843 \epsilon^4 + 192 k^2 ( 79 \epsilon^2 - 840 i \omega)+  \\
    &&
    \qquad \qquad - 113{,}264 i \epsilon^2 \omega - 3{,}494{,}400 \omega^2 \bigr) u^2 + \ldots
    \ , \nonumber \\
    a_x^3 &=& a_{x0} - 
     \frac{a_{t0} k + a_{x0} \omega}{{\mathcal P}}
     \frac{\omega}{96} \bigl(26{,}880 k^4 + 843 \epsilon^4 + 192 k^2 (79 \epsilon^2 - 840 i \omega)
    \\
     &&
     \qquad \qquad - 113{,}264 i \epsilon^2 \omega - 3{,}494{,}400 \omega^2 \bigr) u^2 + \ldots
          \nonumber      \ .
\end{eqnarray}
The pole here is again a cubic polynomial in $\omega$:
\begin{eqnarray}
{\mathcal P} &=& - 72{,}800 i \omega^3  + (39{,}760 k^2  + 843 \epsilon^2) \omega^2
+ \frac{5}{6} i  (2688 k^4 + 451 k^2 \epsilon^2 ) \omega \nonumber  \\
&& - \frac{71}{32} k^2 \epsilon^4 - 53 k^4 \epsilon^2 - 280 k^6 \ .
\end{eqnarray}
We consider the roots of the polynomial first in the limit $k \ll \epsilon$:
\begin{eqnarray}
\omega &=& \pm \sqrt{\frac{1}{2} \frac{71}{13{,}488}} \epsilon k - \frac{103{,}535}{947{,}532} i k^2 + \ldots
 \ , \\
\omega &=& - \frac{843}{72{,}800}i \epsilon^2 - \frac{5{,}044{,}459}{15{,}397{,}395} i k^2 + \ldots \ .
\end{eqnarray}
The first pair of poles correspond to second sound propagating in the direction parallel to
the order parameter with a speed consistent with our earlier result (\ref{speedsoundpar}).
The third pole is related to the zero mode that causes a phase transition from the superfluid to
the normal phase.
Next we consider the limit $k \gg \epsilon$:
\begin{eqnarray}
\omega &=& \frac{\pm 11 - 3i}{130} k^2 + \frac{\pm 192{,}553 - 95{,}119i}{26{,}644{,}800} \epsilon^2  
\ , \\
\omega &=& - \frac{ik^2}{2} - \frac{13 i \epsilon^2}{2928} + \ldots \ .
\end{eqnarray}
The two sound poles have evolved into the zero mode poles, while the zero mode pole has evolved
into a diffusive pole.

From the small $u$ expansion, we can read off a large number of Green's functions which we shall
not bother to list.  Similar to the transverse sound case considered above, 
the prefactor $(a_{t0} k + a_{x0} \omega)$ in the expansion means that the
Ward identities (\ref{Wardone}) will be satisfied.  However, there is more structure here.
Note that $ik a_x^2 = 4 a_t^1 - a_{x0} \epsilon u^2$ and $i k a_x^1 = -4 a_t^2$.  
In our hydrodynamic limit at leading order in $\omega$ and $k$, 
these two equations
are the Ward identities (\ref{Wardtwo}) and (\ref{Wardthree}), respectively.

Before moving on, we note that
\be
\lim_{k \to 0} G^{tt}_{33} (0, k) = \frac{2}{g^2} \frac{281}{71} \ ,
\ee
which agrees with eq.\ (\ref{chi}), but $k$ here is parallel rather than transverse to the order parameter.

\subsection{Transverse diffusive mode}

In addition to the sound mode found above, in the limit $k \ll \epsilon$, we expect
to find a diffusive mode in the current-current correlator.  We begin with the slightly simpler
case of a mode polarized transverse to the order parameter but propagating parallel to it, and
follow in the next section with a mode polarized longitudinal to the order parameter but
propagating transversely.
Thus first we look for fluctuations in $\delta A_y^a(u,t,x)$ and any other modes that couple to it.
A self-consistent set of fluctuations to consider is
\be
\delta A_y^a(u,t,x) = a_y^a(u) e^{-i \omega t + i k x} \ , 
\ee
where $a=1,2,3$.

This set of fluctuating modes gives rise to the three differential equations at linear order:
\begin{eqnarray}
\cald_x a_y^3 &=& \frac{ (k^2 + W^2) f - \omega^2}{f^2} a_y^3 - \frac{2i W}{f}a_y^2 
\ , \\
\cald_x a_y^2 &=& \frac{(k^2 + W^2) f - \omega^2 - \Phi^2}{f^2} a_y^2 + \frac{2i W}{f} a_y^3 
- \frac{2 i \omega \Phi}{f^2} a_y^1 
\ , \\
\cald_x a_y^1 &=& \frac{k^2 f - \omega^2 - \Phi^2}{f^2} a_y^1 + \frac{2i \omega \Phi}{f^2} a_y^2 
 \ .
\end{eqnarray}

As before, we solve this set of equations perturbatively in the limit $\omega \sim k^2 \sim \epsilon^2$.
The small $u$ expansion of the solutions, from which we may read off the Green's functions, takes the
form:
\begin{eqnarray}
a_y^1 &=&  \frac{a_{y0}}{{\mathcal P}} 22 k \epsilon \omega u^2 + \ldots
\ , \\
a_y^2 &=& \frac{a_{y0}}{{\mathcal P}}2  k \epsilon (2 i k^2 + 3 \omega) u^2 + \ldots
\ , \\
a_y^3 &=& a_{y0} -\frac{a_{y0}}{{\mathcal P}} \frac{1}{3360} \Bigl(
-1680 k^6 + 12 k^4 ( 29 \epsilon^2 + 700 i \omega) -
6 k^2 ( \epsilon^4 + 28 i \epsilon^2 \omega - 10{,}780 \omega^2) + \nonumber \\
&&
\qquad + \omega (9 i \epsilon^4 + 4766 \epsilon^2 \omega - 109{,}200 i \omega^2) \Bigr) u^2
+ \frac{1}{2} a_{y0} k^2 u^2 \ln u + \ldots
\ .
\end{eqnarray}
The poles at leading order in this perturbative expansion 
come from a quadratic polynomial in $\omega$:
\be
{\mathcal P} =  65 \omega^2 +\frac{3i(140 k^2 + 3 \epsilon^2)}{70} \omega 
- \frac{ (70 k^2 + 3 \epsilon^2)k^2}{35} \ .
\ee
As before, we consider the roots of this polynomial in two limits.  First we consider $k \ll \epsilon$, 
in which case we find
\begin{eqnarray}
\omega &=& - \frac{9i}{4550} \epsilon^2 + \frac{112 i}{195} k^2 + \ldots \ , \\
\omega &=& - \frac{2ik^2}{3} + \ldots \ .
\end{eqnarray}
The first pole is associated with the zero mode that causes the phase transition from the superfluid
phase to the normal phase while the second pole comes from a diffusive mode of the charge density.

Next, we consider the limit $k \gg \epsilon$ where we recover the zero modes of the normal phase, 
\be
\omega = \frac{\pm 11 - 3i}{65}k^2 + \frac{ \pm 33 - 9i}{9100} \epsilon^2 + \ldots \ .
\ee
At leading order, the location of the pole is the same as that of eq.\ (\ref{zmodetsound}).
However, the subleading order $\epsilon^2$ corrections are different.

\subsection{Longitudinal diffusive mode}

We continue the discussion by looking at modes polarized longitudinal to the order parameter
but propagating transversely.  We consider fluctuations $\delta A_x^3(u,t,y)$ and all others
coupled to it:
\begin{eqnarray}
\delta A_x^3(u,t,y) &=& a_x^3(u) e^{-i \omega t + i k y} \ , \nonumber \\
\delta A_t^a(u,t,y) &=& a_t^a(u) e^{-i \omega t + i k y} \ , \\
\delta A_y^b(u,t,y) &=& a_y^b(u) e^{-i \omega t + i k y} \ , \nonumber
\end{eqnarray}
where $a, b = 1,2$. 
This set of fluctuations obeys the 
five second order equations and two first order constraints:
\begin{eqnarray}
\cald_t a_t^1 &=& \frac{1}{f} \left( -W \Phi  a_x^3-i k \Phi  a_y^2+k^2 a_t^1+k \omega  a_y^1  \right)
\ , \\
\cald_t a_t^2 &=& \frac{1}{f} \left( i W \omega  a_x^3 +\left (k^2 + W^2 \right) a_t^2 + 
 k \omega  a_y^2 + i k \Phi  a_y^1
\right)
\ , \\
\cald_x a_y^1 &=& \frac{1}{f^2} \left( i k \Phi  a_t^2+2 i \Phi  \omega  a_y^2-k \omega  a_t-\left(\Phi ^2+\omega ^2\right) a_y \right)
\ , \\
\cald_x a_y^2 &=& \frac{1}{f^2} \left(-i f k W a_x^3 - 
 k \omega  a_t^2 + \left (f W^2 - \Phi ^2 - \omega ^2 \right) a_y^2 - i k \Phi  a_t - 2 i \Phi  \omega  a_y
\right)
\ , \\
\cald_x a_x^3 &=& \frac{1}{f^2} \left(\left(f k^2- \omega^2 \right) a_x^3+i W \omega  a_t^2+i f k W a_y^2+W \Phi  a_t^1
\right)
\ , \\
0 &=&  -\Phi ' a_t^2+i \omega  \partial_u a_t^1+\Phi  \partial_u a_t^2+i f k \partial_u a_y^1 \ , \\
0 &=& f W' a_x^3+ \Phi ' a_t^1 -\Phi  \partial_u a_t^1+i \omega  \partial_u a_t^2-f W \partial_u a_x^3+i f k \partial_u a_y^2 \ .
\end{eqnarray}
At the horizon, there are two pure gauge solutions, three ingoing solutions, and three outgoing solutions.  We discard the outgoing solutions and use the remaining degrees of freedom to choose the boundary
values of the five fluctuations.  In particular, we set the boundary values of all the fluctuations to zero
save for $a_x^3$, which we set to $a_{x0}$.  The near boundary expansion of the solution takes
the form:
\begin{eqnarray}
a_t^1 &=& \frac{a_{x0}}{{\mathcal P}} \frac{3 k^2 \epsilon}{4} (k^2 - 3 i \omega)u^2  +  a_{x0} \frac{\epsilon}{4} u^2 
+ \ldots
\ , \\
a_t^2 &=& \frac{a_{x0}}{{\mathcal P}} \frac{33 i k^2 \omega \epsilon}{4}
u^2 + \ldots
\ , \\
a_y^1 &=& - \frac{a_{x0}}{{\mathcal P}} 33 k \epsilon \omega \, u^2 + \ldots
\ , \\
a_y^2 &=& - \frac{a_{x0}}{{\mathcal P}} 3 i k \epsilon (k^2 - 3 i \omega) u^2 + \ldots
\ , \\
  a_x^3 &=& a_{x0} + \frac{a_{x0}}{{\mathcal P}} \frac{1}{3360}
\Bigl( 840 k^6 - 16 k^4 (13\epsilon^2 + 420 i \omega) -
 \omega (\epsilon^2 -48 i \omega) (9 i \epsilon^2 + 4550 \omega) \nonumber \\
&& + 3k^2 ( \epsilon^4 + 125 i \epsilon^2 \omega - 39{,}760 \omega^2) \Bigr)u^2
+ \frac{1}{2} a_{x0} k^2 u^2 \ln u + \ldots
\ .
\end{eqnarray}
The pole, similar to the case considered previously, is a quadratic polynomial in $\omega$:
\be
{\mathcal P} =  130 \omega^2 +\left( 6 i k^2 + \frac{9 i \epsilon^2}{35} \right) \omega 
- \frac{3}{35} k^2 \epsilon^2 - k^4 \ .
\ee
In the limit $k \ll \epsilon$ we find a zero mode and a diffusive mode:
\begin{eqnarray}
\omega &=& - \frac{9i}{4550} \epsilon^2 + \frac{56 i}{195} k^2 + \ldots \ , \\
\omega &=& - \frac{ik^2}{3} + \ldots \ .
\end{eqnarray}
In the opposite limit, we find two zero modes:
\be
\omega = \frac{\pm 11 - 3i}{130}k^2 + \frac{ \pm 33 - 9i}{9100} \epsilon^2 + \ldots \ .
\ee

The structure of the small $u$ expansion of the gauge fields is again related to the Ward identities.
We see that $ik a_y^2 = 4 a_t^1 - a_{x0} \epsilon u^2$ and $ik a_y^1 = - 4 a_t^2$, which are
restatements of the Ward identities (\ref{Wardtwo}) and (\ref{Wardthree}), respectively.

We have thus far considered the current-current correlation functions in five special cases.
While the results are simpler, our presentation has the disadvantage of 
obscuring the relationship between the various cases.  We make some remarks about 
and provide some results for the 
general case in Appendix \ref{sec:generalcase}.

\subsection{Conductivity and London Equations}
\label{sec:conductivities}

In this section, we begin by studying the response of the system to a homogeneous, time
dependent electric field, $\delta A_j^3 \sim e^{-i \omega t}$, and end with a discussion 
of the London equations.  
A homogeneous electric field should produce a current in the system via Ohm's Law.
To investigate the conductivity in this long wavelength limit,
we set $k= 0$ for the two-point functions computed above.  

The case of an electric field orthogonal to the order parameter is simple; 
a current and nothing more is produced.
From the pure transverse mode and eq.\ (\ref{puretransverseG}), we have
\be
G_{33}^{yy}(\omega) = \frac{2}{g^2} \left(- \frac{\epsilon^2}{48} + i \left(\frac{1}{2}  - \frac{\epsilon^2}{48} \right) \omega
+ c \, \omega^2 \right) + \ldots \ .
\ee
Reassuringly, this result agrees with the $k\to 0$ limit of the Green's functions associated to
transverse sound propagation and the transverse diffusive mode.

For an electric field parallel to the order parameter, the physics is richer.  We find a current
in the $x$ direction but also oscillating (or precessing) charge densities associated with the one and two
isospin directions:
\begin{eqnarray}
 a_x^3 &=& a_{x0} + a_{x0} \left( - \frac{\epsilon^2}{96} + i \left( \frac{1}{2} + \frac{\epsilon^2}{288} \right) \omega \right) \rv^2 + \ldots \ , \\
a_t^1 &=& a_{x0} \frac{\epsilon}{4} \rv^2 + \ldots \ , \\
a_t^2 &=& -a_{x0} \frac{i \epsilon\omega}{16} \rv^2 + \ldots \ .
\end{eqnarray}
This near boundary expansion agrees with the $k\to 0$ limit of the expansions for longitudinal sound and diffusion considered above.  The associated Green's functions are
\begin{eqnarray}
G^{xx}_{33}(\omega) &=& \frac{2}{g^2} \left( - \frac{\epsilon^2}{96} + i \left( \frac{1}{2} + \frac{\epsilon^2}{288} \right) \omega \right) + \ldots
\ , \\
G^{tx}_{13}(\omega) &=&  -\frac{2}{g^2}\frac{\epsilon}{4} + \ldots
\ , \\
G^{tx}_{23}(\omega) &=& \frac{2}{g^2}\frac{i \epsilon\omega}{16} + \ldots
\ .
\end{eqnarray}

\begin{figure}
  \centerline{\includegraphics[width=5in]{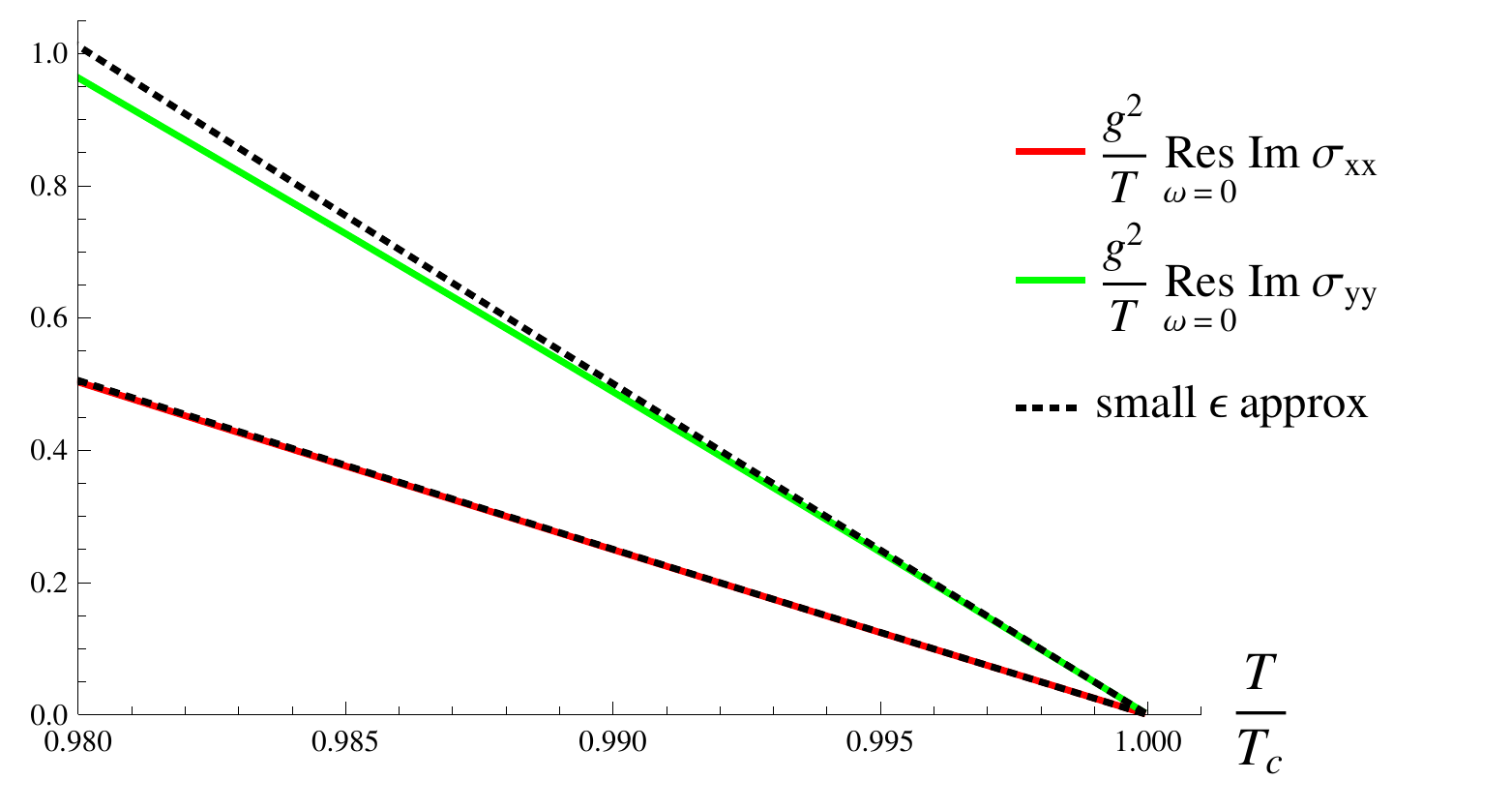}}
  \caption{Plots of numerical results for ${g^2 \over T} {\rm Res}_{\omega = 0} {\rm Im} \sigma_{xx}$ and ${g^2 \over T} {\rm Res}_{\omega = 0} {\rm Im} \sigma_{xx}$ as functions of temperature (solid lines), as well as analytical approximations at small $\epsilon$ (dotted lines) given by equations \eqref{GotResidues}. }
  \label{NumericalResidues}
\end{figure}

Identifying the electric field $E_j = i \omega \, \delta A_j$ and recalling Ohm's Law, 
the conductivities are related via eq.\ (\ref{twopointformula}) to the retarded Green's functions, 
\be
\label{SigmaFromG}
\sigma_{xx}(\omega) = \frac{G^{xx}_{33}(\omega)}{i \omega} 
\qquad 
\mbox{and}
\qquad
\sigma_{yy}(\omega) = \frac{G^{yy}_{33}(\omega)}{i \omega} \ .
\ee
The terms proportional to $\epsilon^2$ in $G^{xx}_{33}$ and $G^{yy}_{33}$ thus produce a pole
in the imaginary part of the respective conductivities.  As discussed 
in refs.\ \cite{Hartnoll:2008vx, Hartnoll:2008kx}, 
by the Kramers-Kronig relations (or by
properly regularizing the pole) there must be a delta function in the real part of the conductivity,
indicating the material loses all resistance to DC currents and suggesting the phase transition
is to a superconducting state.  While in refs.\ \cite{Hartnoll:2008vx, Hartnoll:2008kx}, the pole was seen only numerically,
here we can calculate the strength of the pole analytically close the phase transition. 
 Its residue is given by
 \begin{equation}
  \label{GotResidues}
  {\rm Res}_{\omega = 0} \sigma_{xx} = {2 \over g^2} {i\epsilon^2 \over 96} + \ldots
  \qquad 
  {\rm Res}_{\omega = 0} \sigma_{yy} = {2 \over g^2} {i\epsilon^2 \over 48} + \ldots \,.
 \end{equation}
 In Figure~\ref{NumericalResidues} we show a comparison between numerical computations of the residues of the poles at $\omega = 0$ in $\sigma_{xx}$ and $\sigma_{yy}$, along with the analytic approximation \eqref{GotResidues} close to $T = T_c$.

In the Drude model for an ideal metal, the conductivity takes the form $\sigma = i \rho / m \omega$ where $\rho$ is the charge density and $m$ is the mass of the charge carrier.  In the superconductivity literature (see for example \cite{Tinkham}), the pole in the imaginary part of the conductivity is thus often related to a superfluid density.  Because our system is not rotationally symmetric, the density to mass ratio 
defined in this way will depend on the orientation of the superfluid velocity with respect to the order parameter.  The proper way to interpret this situation is probably that a suitably defined effective mass of the superfluid depends on the direction of propagation.

An important observation is that in our system, the $\omega \to 0$ and $k \to 0$ limits of the Green's functions commute.  The residue of the pole in the conductivity is related to the long wavelength limit of the current-current correlation function in the following way:
\be
 i \, {\rm Res}_{\omega = 0} \sigma_{jj} = \lim_{\omega \to 0} \lim_{k \to 0} G^{jj}_{33}(\omega, k) \ .
\ee
The limit in the opposite order is related to a thermodynamic susceptibility:
\be
\lim_{k_y \to 0} G^{xx}_{33} (0, k_y) = \frac{\partial^2 P}{\partial v_\parallel^2} \qquad
\mbox{and}
\qquad
\lim_{k_x \to 0} G^{yy}_{33} (0, k_x) = \frac{\partial^2 P}{\partial v_\perp^2} 
\ ,
\ee
where $v_\parallel$ and $v_\perp$ are superfluid velocities.\footnote{%
  Note that to produce a perturbing magnetic field, we require a $k$ that is transverse to the polarization of the current-current correlation function.
  A perturbation of the form $\delta A_x^3 \sim e^{i k x}$ or $\delta A_y^3 \sim e^{i k y}$ 
  is gauge equivalent to zero and does not   
  produce a response from the system.  The Green's function in this limit vanishes.
 }
It follows from eqs.\ \eqref{Ayperp} and \eqref{Aypar} that 
\be
\frac{\partial^2 P}{\partial v_\parallel^2}=
\frac{\partial j_\parallel}{\partial v_\parallel} = -\frac{2}{g^2}\frac{\epsilon^2}{96} 
\qquad
\mbox{while}
\qquad
\frac{\partial^2 P}{\partial v_\perp^2} = \frac{\partial j_\perp}{\partial v_\perp} = -\frac{2}{g^2}\frac{\epsilon^2}{48}  \ .
\ee
When combined with eq.\ \eqref{SigmaFromG}, these results confirm eq.\ \eqref{GotResidues}.
To see how these limits commute in greater detail, the reader is referred to Appendix \ref{sec:generalcase} and 
eqs.~\eqref{G33ttInviscid}--\eqref{G33xyInviscid} that give the general Green's functions in the inviscid
limit $k_x , k_y \ll \epsilon$.  

As emphasized in this context in ref.\ \cite{Hartnoll:2008kx}, that the limits commute implies the system
really does become a superconductor below $T_c$.  Given that the limits commute, 
the system obeys a London type equation for small $k$ and $\omega$:
\be
\langle j_x^3 \rangle \approx -\frac{2}{g^2} \frac{\epsilon^2}{96} {\mathcal A}_x^3 \qquad
\mbox{and}
\qquad
\langle j_y^3 \rangle \approx -\frac{2}{g^2} \frac{\epsilon^2}{48} {\mathcal A}_y^3 \ .
\ee 
If we now imagine the U(1) subgroup generated by $T^3 \in \mathfrak{su}(2)$ is weakly gauged,
these London equations imply not only infinite DC conductivity but also a Meissner effect with 
London penetration depths that scale 
as $\lambda_\perp \sim \lambda_\parallel \sim 1/\epsilon \sim (T_c - T)^{-1/2}$.

\section{Discussion}

One of the nicest features of our results is their analytic nature.  We were able to confirm a number of previous numeric observations \cite{Hartnoll:2008vx, Gubser:2008zu,GubserPufu, Herzog:2008he, Basu:2008st, Horowitz:2008bn} 
of this superfluid phase transition.  In particular, we 
saw explicitly that the phase transition was second order; the difference in free energy between the phases scaled as $(T_c-T)^2$ below the phase transition.  
We saw the order parameter grew as $\epsilon \sim (T_c - T)^{1/2}$ below $T_c$ and thus has
a mean field critical exponent.  We calculated the speed of second sound near the phase transition
and observed that it vanished linearly with the reduced temperature $c_\perp \sim c_\parallel \sim (T_c-T)$.  We looked at the pole at $\omega = 0$ in the imaginary part of the conductivity and saw the same scaling, 
$\sigma \sim i(T_c - T)/\omega$, that had been observed numerically in a related model \cite{Hartnoll:2008vx} and confirmed that the London penetration depth scales as $\lambda \sim 1/(T_c-T)^{1/2}$.  
This laundry list of scalings (or critical exponents) is the same observed in the mean-field Landau-Ginzburg model of a superconductor.

\begin{figure}
  \centerline{\includegraphics[width=5in]{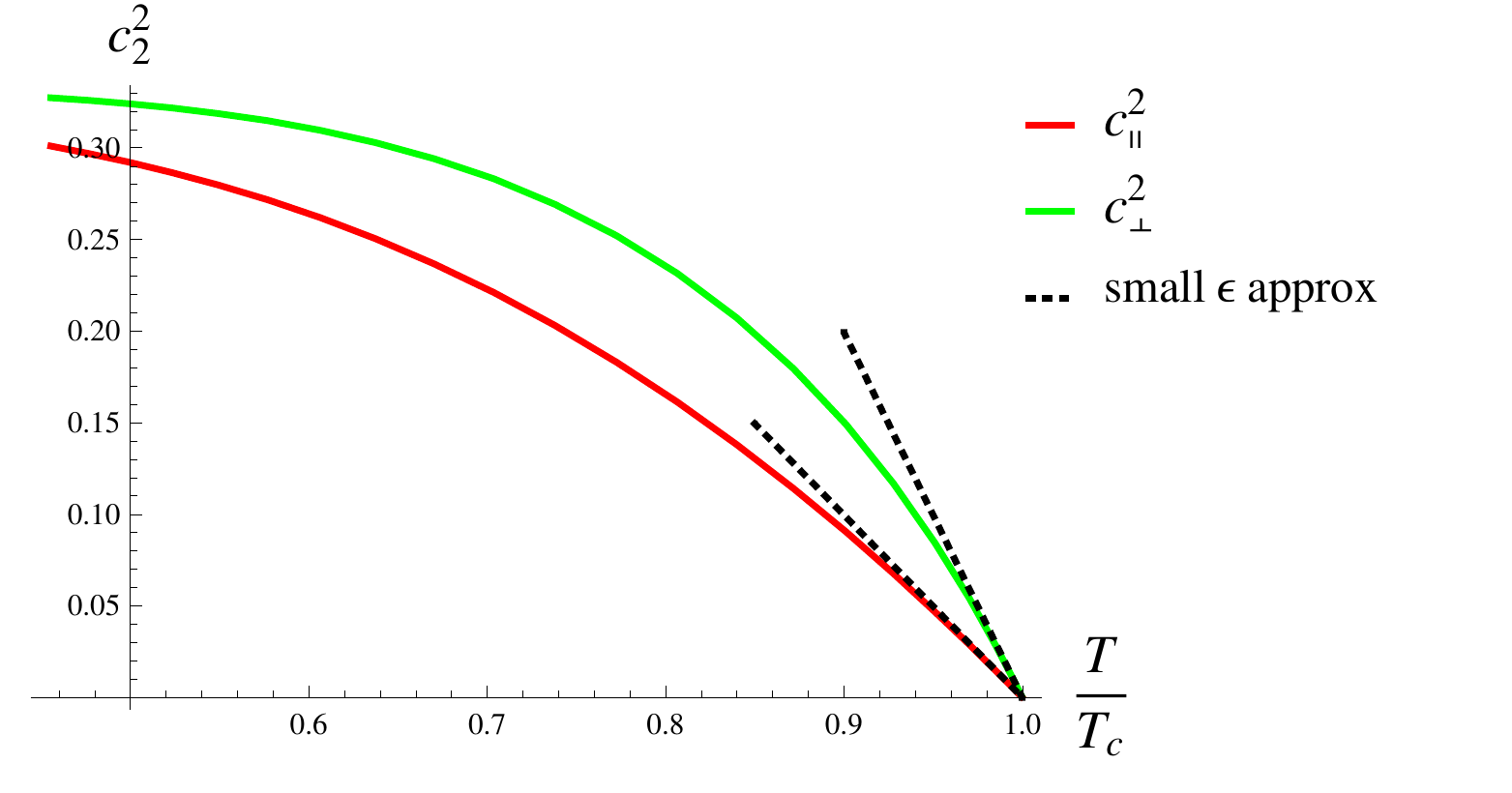}}
  \caption{The squared speeds of second sound $c_\parallel^2$ and $c_\perp^2$ as functions of the reduced temperature $T/T_c$ (solid lines) as well as analytical approximations given by eqs.~\eqref{speedsoundperp} and \eqref{speedsoundpar} close to $T_c$ (dotted lines).
  }
  \label{c2Plot}
\end{figure}

Close to $T_c$, eqs.~\eqref{speedsoundperp} and \eqref{speedsoundpar} indicate that $c_\perp^2 = 2 c_\parallel^2$, so one may wonder whether such a formula is valid away from $T_c$ as well.  Numerical evaluations show that this is not the case:  see Figure~\ref{c2Plot}.  At small temperatures, $c_\perp^2$ approaches $1/3$.  Our numerical evaluations are not sufficiently reliable at small temperatures to see whether $c_\parallel^2$ has the same limit.\footnote{%
 In the case of a scalar order parameter and a phase transition that does not break rotational symmetry,
 we expect the speed of second sound to approach $(d-1)^{-1/2}$ as $T \to 0$.  This limit follows from
 eq.\ (\ref{speedsecondsound}) and two observations: 1) At $T=0$ the Lorentz
 symmetry breaking due to the temperature disappears and the pressure can depend on $\mu$
 and $v$ only as $P(\mu^2 - v^2)$.  2) By dimensional analysis, when $T=v=0$, $P \sim \mu^{d}$. 
 We would like to thank Amos Yarom for discussion on this point.
}

There are some other results in this paper that are worth emphasizing.  For superfluid velocities that are not too large, 
we were able to determine analytically
the critical line in the temperature-superfluid velocity plane separating the normal phase from the superfluid phase.  We calculated a large number of current-current correlation functions in the hydrodynamic limit and verified that they satisfied the non-abelian Ward identities.  
We also investigated how the hydrodynamic poles in these correlation functions move around as a function of $k$ and $\epsilon$.

Optimistically, we hope that someday this system will be more than a toy model.  The introduction described possible similarities of this system to helium-$3$ and $p$-wave superconductors.  Here we add a speculation about a possible connection with QCD\@.  The SU(2) global symmetry of our model could be thought of as the residual approximate isospin symmetry of QCD at low energies and our chemical potential an isospin chemical potential.  The phase structure of QCD at non-zero isospin chemical potential has been discussed by ref.\ \cite{SonStephanov}.  Alas, there is no persistent current in the stable phases they discuss.

In the future, there are some questions that we would like to address.
While we focused on current-current correlation functions in the third isospin direction in this paper,  
it would be good to study the full set of Green's functions more carefully.  Because of the characteristic magnetic properties of superconductors, it would be interesting to investigate the dependence of the correlation functions on an external magnetic field.

Another interesting direction to pursue is the connection between this work and the membrane paradigm \cite{membrane} where the horizon of a black hole, rather than the boundary of an
asymptotically anti-de Sitter space, is thought of as a fluid.  
Related to this direction is the observation of 
ref.\ \cite{Gubser:2008zu} that the fraction of the total charge density outside the black hole horizon scales as $T_c - T$ close to $T_c$, suggesting that this quantity might be related to the superfluid density.  We would like to know if this analogy can be made more precise.

\section*{Acknowledgments}
We would like to thank Steve Gubser, Duncan Haldane, David Huse, Nikola Kamburov, Igor Klebanov, and Amos Yarom for discussion.  The work of CPH was supported in part by the US NSF under Grant No.\ PHY-0756966.  SSP was supported in part by the NSF under award number~PHY-0652782.

\appendix

\section{Higher zero modes}
\label{app:higher_zero_modes}

One can also find other analytical solutions to \eqref{A3t0eqandA1x0eq} when the chemical potential is $\mu = 4k$, where $k$ is an integer, corresponding to higher zero modes of $A_x^1$.  In fact, numerical exploration shows that these are the only values of $\mu$ for which zero modes occur.  The corresponding backgrounds are probably unstable, because they have higher free energy.  

When $\mu = 4 k$, the corresponding zero mode can be written as
 \begin{equation}
  \label{HigherZeroModes}
  A_x^1 = \epsilon \frac{u^2 P_k(u^2)}{(1 + u^2)^{2 k}} \ ,
 \end{equation}
where $\epsilon$ is an arbitrary constant and $P_k(u^2)$ is a polynomial of degree $2(k-1)$ in $u^2$.  It is not hard to find a recursion formula for the coefficients of $P_k(x)$.  Writing
 \begin{equation}
  \label{PkExpansion}
  P_k(x) = \sum_n a_n^{(k)} x^n
 \end{equation}
and using equation \eqref{A3t0eqandA1x0eq}, it follows that
 \begin{equation}
  \label{RecursionFormula}
  n (n + 1) a_n^{(k)} + 4 k (k-n) a_{n-1}^{(k)} 
    - \left[4 k^2 - 2 k(2n - 1) + n(n-1) \right] a_{n-2}^{(k)} = 0 \ .
 \end{equation}
 The boundary conditions needed to solve this recursion relation can be taken to be $a_{-1}^{(k)} = 0$ and $a_{0}^{(k)} = 1$, the latter representing the normalization of the series.  One can check that for each integer value of $k$ the series terminates at order $n = 2k$, the last non-zero coefficient being $a_{2 (k-1)}$.  In table~\ref{PkTable} we provide analytic expressions for the first few $P_k(u^2)$.
 
 \begin{table}[hbdp]
   \caption{Formulas for $P_k(u^2)$ with $k \leq 5$.}
   \begin{center}
    \begin{tabular}{c|l}
      $k$ & $P_k(u^2)$ \\
      \hline
      $1$ & $1$ \\
      $2$ & $1 - 4 u^2 + u^4$ \\
      $3$ & $1 - 12 u^2 + \frac{82}{3} u^4 - 12 u^6 + u^8$ \\
      $4$ & $1 - 24 u^2 + 135 u^4 -240 u^6 + 135 u^8 -24 u^{10} + u^{12}$ \\
      $5$ & $1 - 40 u^2  + 412 u^4 - 1560 u^6 
        + \frac{12126}{5} u^8 - 1560 u^{10} + 412 u^{12} - 40 u^{14} + u^{16}$
    \end{tabular} 
    \end{center}
   \label{PkTable}
  \end{table}%

\section{The General Case}
\label{sec:generalcase}

Consider a general fluctuation
of the form 
$\delta A^a_\mu = a^a_\mu(u) e^{-i \omega t + i k_x x + i k_y y}$ for $a = 1,2,3$ and $\mu= t,x,y$,
in the limit where $\omega \sim k_x \epsilon \sim k_y \epsilon$.  Solving the system
of differential equations in the hydrodynamic limit leads to a set of Green's functions with a 
pole that is a fifth order polynomial in $\omega$.  Three of the roots of this polynomial
correspond to the roots of the third order polynomial we found for the sound modes.
Two of the roots of this polynomial correspond to the roots of the second order polynomial
we found for the diffusive modes.  

In other words, the six modes we found in the longitudinal and transverse sound cases are actually different limits ($k_y \to 0$ and $k_x \to 0$, respectively) of three of the roots of this fifth order polynomial in $\omega$.  Similarly, the four modes we found in the longitudinal and transverse diffusion cases are different limits of two of the roots of this fifth order polynomial.

In the limit $k_x, k_y \ll \epsilon$, we find that the roots are
\begin{eqnarray}
\omega &=& \pm \epsilon \sqrt{\frac{71}{13{,}488}}  \sqrt{\frac{1}{2} k_x^2 + k_y^2}
- \frac{i(103{,}535 \, k_x^2 + 147{,}217 \, k_y^2)}{947{,}532} + \ldots \ , \\
\omega &=& - \frac{i}{3} (2 k_x^2 + k_y^2) + \ldots \ , \\
\omega &=& - \frac{9 i \epsilon^2}{4550} + \frac{56}{195}i (2 k_x^2 + k_y^2) + \ldots \ , \\
\omega &=& - \frac{843 i \epsilon^2}{72{,}800} - \frac{7i(720{,}637 \, k_x^2 + 619{,}349 \, k_y^2)}{15{,}397{,}395} + \ldots \ . 
\end{eqnarray}
In the limit $k_x, k_y \gg \epsilon$, we find in contrast that
\begin{eqnarray}
\omega &=& \left( \pm \frac{11}{65} - \frac{3i}{65} \right) (k_x^2 + k_y^2) +  \\
&& +
\frac{\epsilon^2 }{k_x^2 + k_y^2} \left(
\left( \pm \frac{33}{9100} - \frac{9i}{9100} \right) k_x^2 + \left(
\pm \frac{260{,}803}{26{,}644{,}800} - \frac{131{,}519i}{26{,}644{,}800} \right) k_y^2
\right)
\ , \nonumber \\
\omega &=& \left( \pm \frac{11}{130} - \frac{3i}{130} \right) (k_x^2 + k_y^2) + \\
&&
+
\frac{\epsilon^2 }{k_x^2 + k_y^2} \left(
\left(
\pm \frac{192{,}553}{26{,}644{,}800} - \frac{95{,}119i}{26{,}644{,}800} \right) k_x^2 + \left( \pm \frac{33}{9100} - \frac{9i}{9100} \right) k_y^2 
\right) \ , \nonumber \\
\omega &=& - \frac{i}{2} (k_x^2 + k_y^2) - \frac{ i (13 k_x^2 + 5 k_y^2)}{2928 (k_x^2 + k_y^2)} + \ldots \ .
\end{eqnarray}
The case $k_x, k_y \ll \epsilon$ admits an inviscid limit where the two-point functions of the currents in the third isospin direction become
 \begin{align}
   \label{G33ttInviscid}
  G_{33}^{tt} &= - {2 \over g^2} {\epsilon^2 \over 96} 
    {k_x^2 + 2 k_y^2 \over \omega^2 - c_\parallel^2 (k_x^2 + 2 k_y^2)} \ ,\\
  G_{33}^{tx} &= -{2 \over g^2} {\epsilon^2 \over 96}
    {\omega k_x \over \omega^2 - c_\parallel^2 (k_x^2 + 2 k_y^2)} = G_{33}^{xt} \ ,\\
  G_{33}^{ty} &= -{2 \over g^2} {\epsilon^2 \over 96}
    {2 \omega k_y \over \omega^2 - c_\parallel^2 (k_x^2 + 2 k_y^2)} = G_{33}^{yt} \ ,\\ 
  G_{33}^{xx} &= -{2 \over g^2} {\epsilon^2 \over 96}
    {\omega^2 - 2 c_\parallel^2 k_y^2 \over \omega^2 - c_\parallel^2 (k_x^2 + 2 k_y^2)} \ ,\\  
  G_{33}^{yy} &= -{2 \over g^2} {\epsilon^2 \over 96}
    {2 (\omega^2 - c_\parallel^2 k_x^2) \over \omega^2 - c_\parallel^2 (k_x^2 + 2 k_y^2)} \ ,\\ 
   \label{G33xyInviscid}     
  G_{33}^{xy} &= -{2 \over g^2} {\epsilon^2 \over 96}
    {2 c_\parallel^2 k_x k_y \over \omega^2 - c_\parallel^2 (k_x^2 + 2 k_y^2)} = G_{33}^{yx} \ .
 \end{align}
From eqs.~\eqref{G33ttInviscid}--\eqref{G33xyInviscid} one can easily see that the limits $\omega \to 0$ and $k \to 0$ commute, implying that there is a Meissner effect;  see Section~\ref{sec:conductivities}.

Another interesting limit to take of this fifth order polynomial in $\omega$ is the limit $\omega \to 0$:
\begin{equation}
\begin{split}
{\mathcal P} \sim\  &(k_x^2+k_y^2)^2 \left( k_x^2 + k_y^2 + \frac{\epsilon^2}{16} \right) \\
&\times \left(k_x^2 + k_y^2 + \frac{3\epsilon^2}{70} \frac{k_x^2  + 2 k_y^2 }{k_x^2 + k_y^2} \right)
\left( k_x^2 + k_y^2 + \frac{71 \epsilon^2}{1120} 
\frac{2 k_x^2  + k_y^2}{k_x^2+k_y^2} \right)\ .
\end{split}
\end{equation}
From this limit, one can extract correlation lengths that scale as $1/\epsilon \sim (T_c - T)^{-1/2}$.


\end{document}